\documentclass[a4paper,fleqn,usenatbib]{mnras}

\usepackage[T1]{fontenc}
\usepackage{ae,aecompl}

\usepackage{graphicx}	
\usepackage{amsmath}	
\usepackage{amssymb}	

\usepackage{caption}
\usepackage{subcaption}
\usepackage{mwe}

\usepackage{newtxtext,newtxmath}

\title[Giant Arc]{A Giant Arc on the Sky}

\author[A.M. Lopez et al.]
{Alexia M. Lopez,$^1$\thanks{E-mail: amlopez@uclan.ac.uk}
Roger G. Clowes$^1$
and
Gerard M. Williger$^2$ \\
$^1$ Jeremiah Horrocks Institute, University of Central Lancashire,
Preston PR1 2HE \\
$^2$ Department of Physics and Astronomy, University of Louisville,
Louisville, KY 40292, USA \\
}

\date{Accepted 2022 Xxxxxxx XX. Received 2022 Xxxxxxx XX; in original form
2021 Xxxxxxx XX}

\pubyear{2022}

\begin{document}
\label{firstpage}
\pagerange{\pageref{firstpage}--\pageref{lastpage}}
\maketitle

\begin{abstract}

We present the serendipitous discovery of a `Giant Arc on the Sky' at $z \sim
0.8$. The Giant Arc (GA) spans $\sim 1$~Gpc (proper size, present epoch), and
appears to be almost symmetrical on the sky. It was discovered via
intervening Mg~{\sc II} absorbers in the spectra of background quasars, using
the catalogues of Zhu \& M\'enard. The use of Mg~{\sc II} absorbers
represents a new approach to the investigation of large-scale structures
(LSSs) at redshifts $0.45 \la z \la 2.25$. We present the observational
properties of the GA, and we assess it statistically using methods based on:
(i) single-linkage hierarchical clustering ($\sim 4.5\sigma$); (ii) the
Cuzick-Edwards test ($\sim 3.0\sigma$); and (iii) power spectrum analysis
($\sim 4.8\sigma$). Each of these methods has distinctive attributes and
powers, and we advise considering the evidence from the ensemble.  We discuss
our approaches to mitigating any {\it post-hoc} aspects of analysing
significance after discovery. The overdensity of the GA is $\delta \rho /
\rho \sim 1.3 \pm 0.3$. The GA is the newest and one of the largest of a
steadily accumulating set of very large LSSs that may (cautiously) challenge
the Cosmological Principle, upon which the `standard model' of cosmology is
founded. Conceivably, the GA is the precursor of a structure like the Sloan
Great Wall (but the GA is about twice the size), seen when the Universe was
about half its present age.

\end{abstract}


\section{Introduction}

We are using intervening Mg~{\sc II} absorbers to investigate cosmological
structure at redshifts $0.45 \la z \la 2.25$. Previous work at such redshifts
has generally depended on either: (i) quasars, for which there are accurate
spectroscopic redshifts, or (ii) galaxies (and clusters), for which there are
typically only somewhat less accurate photometric redshifts. With Mg~{\sc II}
absorbers, however, we can obtain accurate spectroscopic redshifts for faint
galaxies, subject, of course, to the limitations that the detected galaxies
will be relatively sparse and that they can be only at the sky coordinates of
the background quasars that are used as probes. These limitations need not,
however, be a restriction on investigating large-scale structure (LSS).

In this way, we have serendipitously discovered a `Giant Arc on the Sky' at
$z \sim 0.8$. The Giant Arc (GA) spans $\sim 1$~Gpc (proper size, present
epoch), and appears to be intriguingly symmetrical on the sky. It is one of
the largest of a steadily accumulating set of very large LSSs that may
(cautiously) challenge the Cosmological Principle (CP), upon which the
`standard model' of cosmology is founded. (Note that the word `challenge' is
not synonymous with `contradict', but it does imply something to be
investigated further.)

\subsection{The largest large-scale structures}

\citet{Yadav2010} gave $\sim 370$~Mpc as an ideal or upper limit to the scale
of homogeneity in the concordance cosmology, beyond which departures from
homogeneity should not be evident. As \citet{Yadav2010} state, above this
scale it should not be possible to distinguish a given point distribution
from a homogeneous distribution. We can therefore take $\sim 370$~Mpc as an
indication of the size (and, incidentally, separation also) beyond which LSSs
becomes cosmologically interesting. We list in Table~\ref{tab:lss_list} some
of the very large LSSs that have been reported in the literature, noting
mainly those of (present-epoch) size $\ga 370$~Mpc. Many of the sizes quoted
in this table will be somewhat uncertain, usually because of uncertainty in
the boundaries, but sometimes because of uncertainty in what is being quoted
in the papers (e.g. cosmological model and parameters, cosmological epoch).
Also, in the lower part of Table~\ref{tab:lss_list}, we list some results
that, while not strictly being instances of LSS, are certainly of interest
for considering the validity of the CP.

\begin {table*}
\flushleft
\caption {Some of the very large LSSs reported in the literature.
The columns are: the name of the LSS; the mean redshift; the reported size
in Mpc (present epoch); and references. The lower part of the table lists
some results, which, while not strictly instances of LSS, are of interest
for the CP.}
\small \renewcommand \arraystretch {0.8}
\newdimen\padwidth
\setbox0=\hbox{\rm0}
\padwidth=0.3\wd0
\catcode`|=\active
\def|{\kern\padwidth}
\newdimen\digitwidth
\setbox0=\hbox{\rm0}
\digitwidth=0.7\wd0
\catcode`!=\active
\def!{\kern1.3\digitwidth}
\begin {tabular} {llll}
\\
Name                              & Mean z      & Size          & References                                                                         \\
                                  &             & Mpc           &                                                                                    \\
\\
HCB Great Wall                    & $\sim 2$    & 2000--3000    & \citet{Horvath2014}; \citet{Horvath2020}$^1$                                       \\
Giant GRB Ring                    & 0.82        & 1720          & \citet{Balazs2015}                                                                 \\
Correlated LQG orientations       & 1.0--1.8    & 1600          & \citet{Friday2022}                                                                 \\
U1.27, Huge-LQG                   & 1.27        & 1240          & \citet{Clowes2013}$^2$                                                             \\
Coherent quasar polarisation$^3$  & 1--2        & 1000          & \citet{Hutsemekers1998}; \citet{Hutsemekers2001}; \citet{Hutsemekers2005}          \\
U1.11                             & 1.11        & !780          & \citet{Clowes2012}                                                                 \\
U1.28, CCLQG                      & 1.28        & !630          & \citet{Clowes1991}; \citet{Clowes2012}                                             \\
Sloan Great Wall                  & 0.073       & !450          & \citet{Gott2005}                                                                   \\
South Pole Wall                   & 0.04        & !420          & \citet{Pomarede2020}                                                               \\
Blazar LSS                        & $\sim 0.35$ & !350          & \citet{Marcha2021}                                                                 \\
Local void                        & $< 0.07$    & !300          & \citet{Keenan2013}; \citet{Whitbourn2016}                                          \\
BOSS Great Wall (BGW)             & 0.47        & !250          & \citet{Lietzen2016}                                                                \\
Great Wall                        & 0.029       & !240          & \citet{Geller1989}                                                                 \\
Saraswati supercluster            & 0.28        & !200          & \citet{Bagchi2017}                                                                 \\
\\
CMB anomalies                     &             &               & \citet{Schwarz2016}                                                                \\
Acceleration anisotropy           &             &               & \citet{Colin2019}                                                                  \\
Cluster anisotropy                &             &               & \citet{Migkas2020}                                                                 \\
Quasar dipole                     & 0.5--2.0    &               & \citet{Secrest2021}                                                                \\
Statistical power, LRGs           & $\sim 0.6$  & 1000          & \citet{Thomas2011}                                                                 \\ 
\\
\end {tabular}
\\
Further references.                                                                                                                                  \\
$^1$ See also: \citet{Christian2020}                                                                                                                 \\
$^2$ See also: \citet{Nadathur2013}; \citet{Marinello2016}; \citet{Hutsemekers2014}                                                                  \\
$^3$ See also: \citet{Marcha2021}                                                                                                                    \\
\label{tab:lss_list}
\end {table*}

\subsection{The Mg~{\sc II} approach} \label{MgII_approach}

The presence of metal-rich intervening absorption lines in the spectra of
quasars reveals foreground gas associated with galaxies.  Specifically, the
prominent and distinctive Mg~{\sc II} doublet feature, which can be seen over
a broad range of redshifts, $0.45 \la z \la 2.25$, is strongly associated
with the low-ionised gas around galaxy haloes, and is therefore an easily
identifiable tracer of galaxies. In particular, Mg~{\sc II} is known to trace
the H~{\sc I} regions indicative of star-formation regions. Mg~{\sc II}
absorbers can be expected to be useful tracers of large-scale structure as
they trace metal-enriched gas associated with galaxies and clusters.

It is generally believed that the strength of the absorption doublet (rest
equivalent widths $W_{r,2796}, W_{r,2803}$) arising in the haloes corresponds
to the properties of the galaxy and galaxy clusters, although there is still
uncertainty about the relative importance of morphology, luminosity, impact
parameter, galaxy inclination etc., and combinations of these, that lead to
the different classes (weak, strong) of absorbers. See for example:
\citet{Lanzetta1990}; \citet{Churchill2000}; \citet{Steidel2002};
\citet{Churchill2005}; \citet{Chen2010}; \citet{Bordoloi2011}.

The roughly spherical haloes hosting strong Mg~{\sc II} systems are
considered to extend to radii $43 \le r \le 88$~kpc \citep{Kacprzak2008}. It
has been suggested by \citet{Steidel1995} and \citet{Churchill2005} that
Mg~{\sc II} absorption with rest-frame equivalent width $W_{r,2796} <
0.3$\,\AA\ occurs predominantly in the outer regions of a halo, whereas
stronger absorption with $W_{r,2796} > 0.3$\,\AA\ occurs predominantly in the
inner regions. Such an association of equivalent width with radius is seen
also in C~{\sc IV} and Lyman-limit absorption systems.  Of course, there is
probably patchy structure so generalisations may be misleading.

\citet{Lee2021} investigate the rate of Mg~{\sc II} absorption in and around
clusters of galaxies. They find that although the detection rate per quasar
is higher inside the clusters, the rate is in fact quite low when considering
the number of galaxies in the clusters --- that is, the galaxy-to-absorber
ratio is lower inside clusters, presumably because the environment within
clusters modifies the galaxy haloes.

\subsubsection{Data sources}

We have constructed our Mg~{\sc II} absorber database from the Zhu \&
M\'enard (Z\&M) catalogues that are publicly available at the website {\small
  https://www.guangtunbenzhu.com/jhu-sdss-metal-absorber-catalog}.  We have
used their DR7 and DR12 `Trimmed' catalogues (not the `Expanded'
versions). DR7 and DR12 indicate that the sources of the quasars that have
been used as background probes are data releases 7 and 12 of the Sloan
Digital Sky Survey (SDSS). The detection of the absorbers and the
construction of the catalogues is described in \citet{Zhu2013} and the above
website.

We paired the Z\&M absorber catalogues on RA, Dec to the `cleaned' quasar
databases DR7QSO \citep{Schneider2010} and DR12Q \citep{Paris2017}. Thus the
absorbers can all be associated subsequently with either DR7QSO or DR12Q.

We removed entries for repeat spectra (see the above website) within the Z\&M
DR12(Q) absorber catalogue, thus avoiding duplication of absorbers. There
were no entries for repeat spectra within the Z\&M DR7(QSO) absorber
catalogue.

When a particular absorber (RA, Dec, $z$) appeared in both the Z\&M DR7(QSO)
and DR12(Q) catalogues, we removed the DR7(QSO) entry, thus giving preference
to DR12(Q) parameters. The final database has 63876 Mg~{\sc II} absorbers.

We also produced a corresponding database of probes from the Z\&M `Quasars
searched' catalogues, similarly restricting them to those that appear in
either DR7QSO or DR12Q. This database has 123351 member quasars.

\subsection{Cosmological model}

The concordance model is adopted for cosmological calculations, with
$\Omega_{T0} = 1$, $\Omega_{M0} = 0.27$, $\Omega_{\Lambda 0} = 0.73$, and
$H_0 = 70$~kms$^{-1}$Mpc$^{-1}$. All sizes given are proper sizes at the
present epoch. (For consistency, we are using the same values for the
cosmological parameters that were used by \citet{Clowes2013}.)

\section{The Giant Arc}

The Giant Arc (GA) is a large, filamentary, crescent-shaped structure that
was discovered serendipitously in Mg~{\sc II} catalogues \citep{Zhu2013}.
The GA extends $\sim 1$~Gpc (proper size, present epoch) in its longest
dimension, at a redshift of $z \sim 0.8$. We now discuss: the discovery of
the GA; observational properties; connectivity and statistical properties;
overdensity and comparisons with other data.

\subsection{Discovery of the Giant Arc} \label{secDiscovery}

\begin{figure}
    \centering
    \includegraphics[scale=0.20]{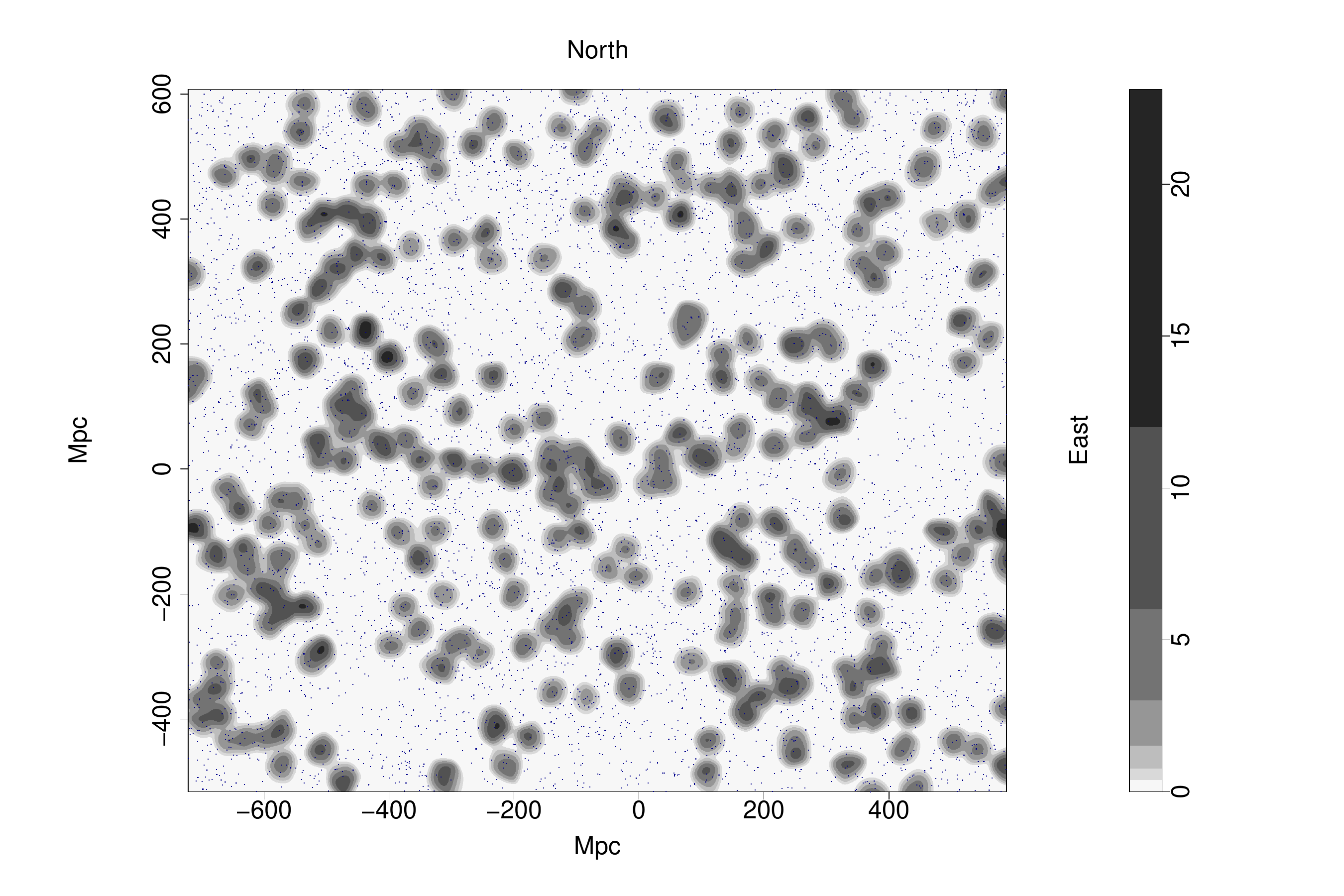}
    \caption{The tangent-plane distribution of Mg~{\sc II} absorbers centred
      in the redshift interval $z=0.802 \pm 0.060$. The grey contours,
      increasing by a factor of two, represent the density distribution of
      the absorbers which have been smoothed using a Gaussian kernel of
      $\sigma = 11$~Mpc, and flat-fielded with respect to the distribution of
      background probes.  The dark blue dots represent the background probes
      (quasars).  The axes are labelled in Mpc, scaled to the present
      epoch. East is towards the right and north is towards the top.  The GA
      runs west-east in the centre of the figure, spanning $\sim 1$ Gpc.}
    \label{fig:The_GA_MgII_image}
\end{figure}

The discovery and the preliminary analysis of the GA were first discussed in
\citet{Lopez2019} and are briefly summarised here. The GA was discovered when
testing the Mg~{\sc II} approach on six small fields containing published
structures (e.g.\ clusters and superclusters) at $z \ga 0.7$.  The
Sunyaev-Zeldovich (SZ) cluster candidate PSZ2~G069.39+68.05
\citep[][subsequently B18]{Burenin2018}, at $z = 0.763$, was one such field:
it indicated a very large LSS extending on the sky, as a dense, long, thin
band of Mg~{\sc II} absorbers, roughly symmetrically to both sides of the
cluster. In \citet{Lopez2019} the central redshift and the redshift interval
of the GA were estimated by stepping through thin redshift slices and
visually inspecting the density and connectivity of the Mg~{\sc II}
absorbers. We have since refined a little the estimates of the central
redshift and the redshift interval --- see Section \ref{subsect:MST} for the
details. In Fig.~\ref{fig:The_GA_MgII_image} the GA can be seen stretching
$\sim 1$ Gpc (proper distance, present epoch) horizontally across the centre
of the field.  Visually, the GA appears densely concentrated and with the
distinctive shape of a giant arc.

The Mg~{\sc II} density images, such as Fig.~\ref{fig:The_GA_MgII_image}
here, and others throughout, are intended to give a useful impression of the
connectivity of the absorbers. The images are constructed by smoothing the 2D
distributions of the Mg~{\sc II} absorbers and the background probes
(quasars) with a Gaussian kernel, with the same smoothing scale for both. In
a process of `flat-fielding' the absorber image is then divided by the
normalised probe image to correct for non-uniformities in the distribution of
the probes on that smoothing scale.  The grey contours in the Mg~{\sc II}
density images increase by a factor of two.  We use tangent-plane
coordinates, scaled, using the central redshift, to present-epoch proper
coordinates in Mpc.

\subsection{Observational properties of the Giant Arc}
\label{sec:observational_properties_GA}

We investigate the observational properties of the GA, in a
visual manner, including: equivalent width (EW) distribution ($W_{r,2796}$);
signal-to-noise ratio (S/N) of the $\lambda 2796$ Mg~{\sc II} line; S/N of
the continuum of the spectra; the $i$~magnitude ($i$) of the probes (background
quasars); and the redshift distribution ($z_{2796}$). Previously, we noted
that EW distribution could be related to the galaxy properties (morphology,
luminosity, impact parameter, galaxy inclination etc.), but these aspects are
still not fully understood. While it may not yet be clear what the EW
distribution within the GA indicates, future studies of EW in Mg~{\sc II}
data should ultimately lead to more understanding of the origins of the GA
and its environment.

The values of the EW are often classed as `strong' or `weak', although there
seems to be no agreement on what defines `strong' and `weak'. For example, in
the literature one might find strong EW variously defined as $W_{r,2796} \ge
0.3$\,\AA, $W_{r,2796} \ge 0.6$\,\AA, and $W_{r,2796} \ge 1.0$\,\AA\ --- see for
example \citet{Churchill2005}, \citet{Dutta2017}, \citet{Evans2013} and
\citet{Williger2002}. We shall follow \citet{Zhu2013} and use the definitions
of strong and weak EWs as $W_{r,2796} \ge 0.6$\,\AA\ and $W_{r,2796} < 0.6$\,\AA\
respectively.

\begin{figure*}
  \centering
   \begin{subfigure}[b]{0.475\textwidth}
        \centering
        \includegraphics[width=\textwidth]{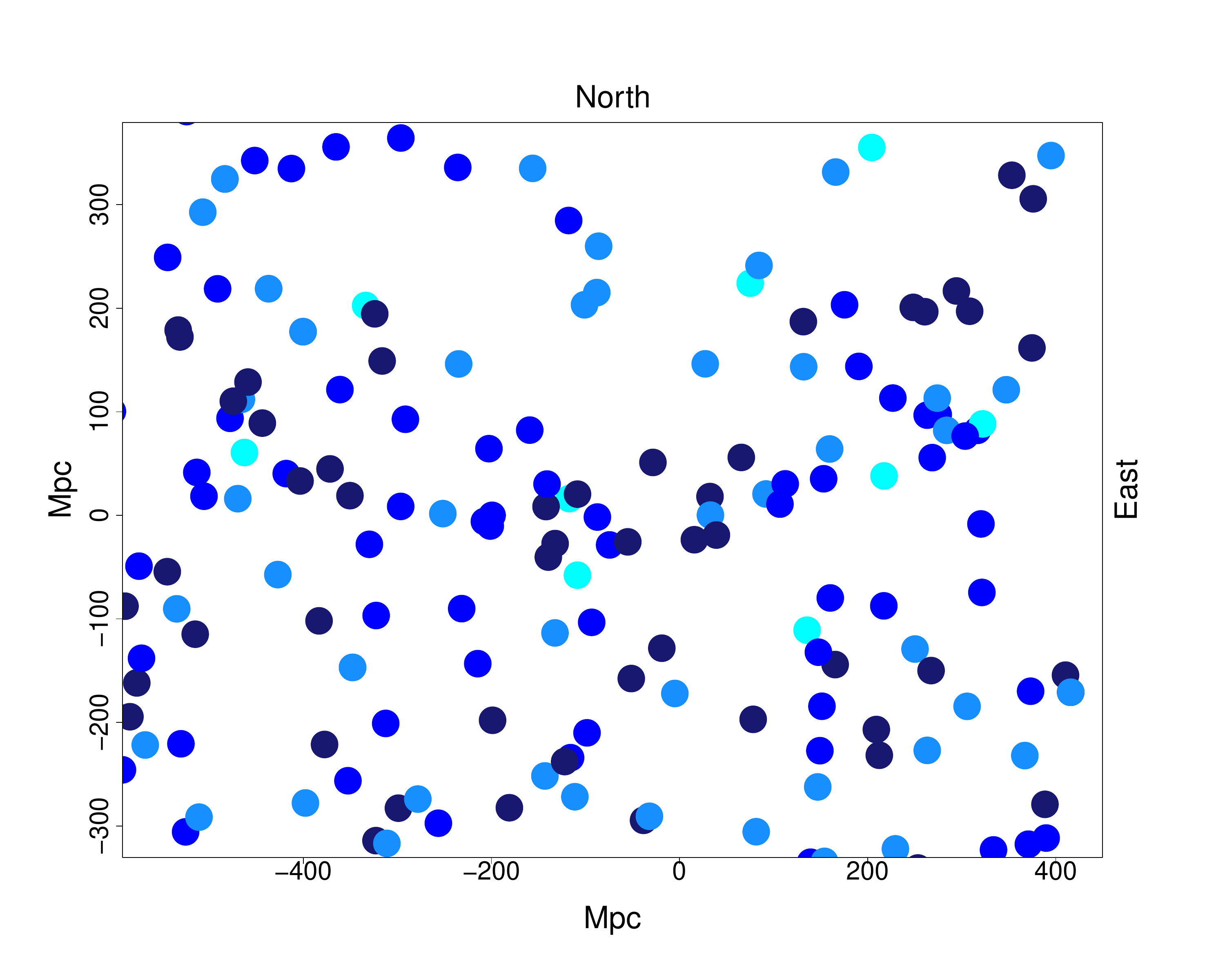}
        \caption {\small The on-sky spatial coordinates of the absorbers in
          the GA and its immediate field, with colour-coding according to the
          four equivalent width (EW) bins of the $\lambda 2796$ Mg~{\sc II}
          line ($W_{r,2796}$). The four EW bins are divided: $0.0, 0.3, 0.6,
          1.0, 10.0$\,\AA, with the lightest shades of blue representing the
          smallest values. There are more strong Mg~{\sc II} absorbers on the
          LHS (lower RA) of the GA than the RHS. Also there appears to be a
          tendency for the strong Mg~{\sc II} absorbers in the GA to clump
          into groups of a few.}
        \label{fig:Rplot_abs_EW_col}
  \end{subfigure}
    \hfill
   \begin{subfigure}[b]{0.475\textwidth}  
        \centering 
        \includegraphics[width=\textwidth]{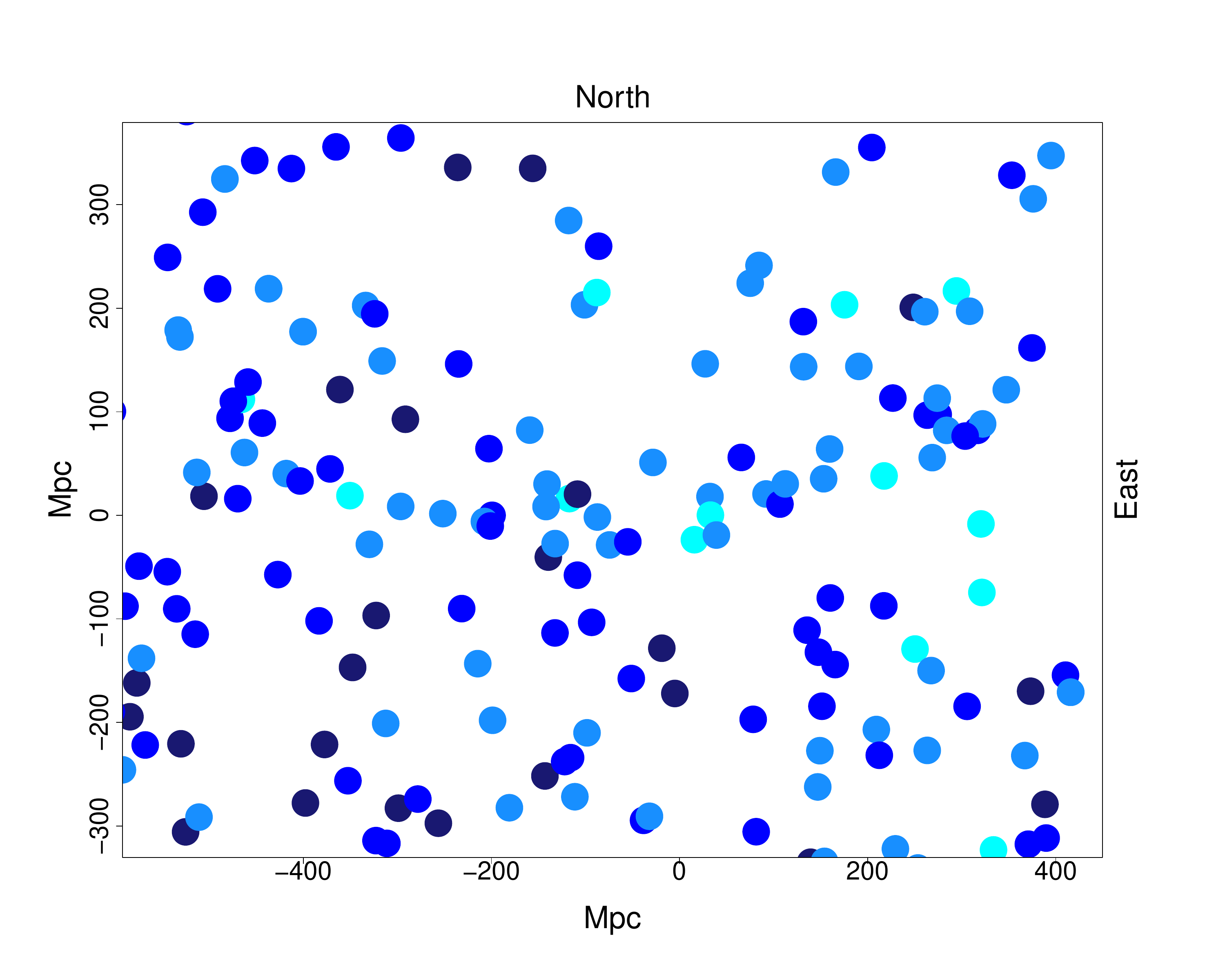}
        \caption {\small The on-sky spatial coordinates of the absorbers in
          the GA and its immediate field, with colour-coding according to the
          four $i$ magnitude bins of the background quasars (probes).  The
          $i$ bins are divided: $16.0, 17.8, 18.7, 19.6, 21.0$, with the
          lightest shades of blue representing the smallest values (this
          means that the lightest shade thus represents the brightest
          quasars). There appear to be more bright quasars on the RHS (higher
          RA) of the GA than the LHS. }
        \label{fig:Rplot_abs_imag_col}
  \end{subfigure}
    \vskip\baselineskip
   \begin{subfigure}[b]{0.475\textwidth}   
        \centering 
        \includegraphics[width=\textwidth]{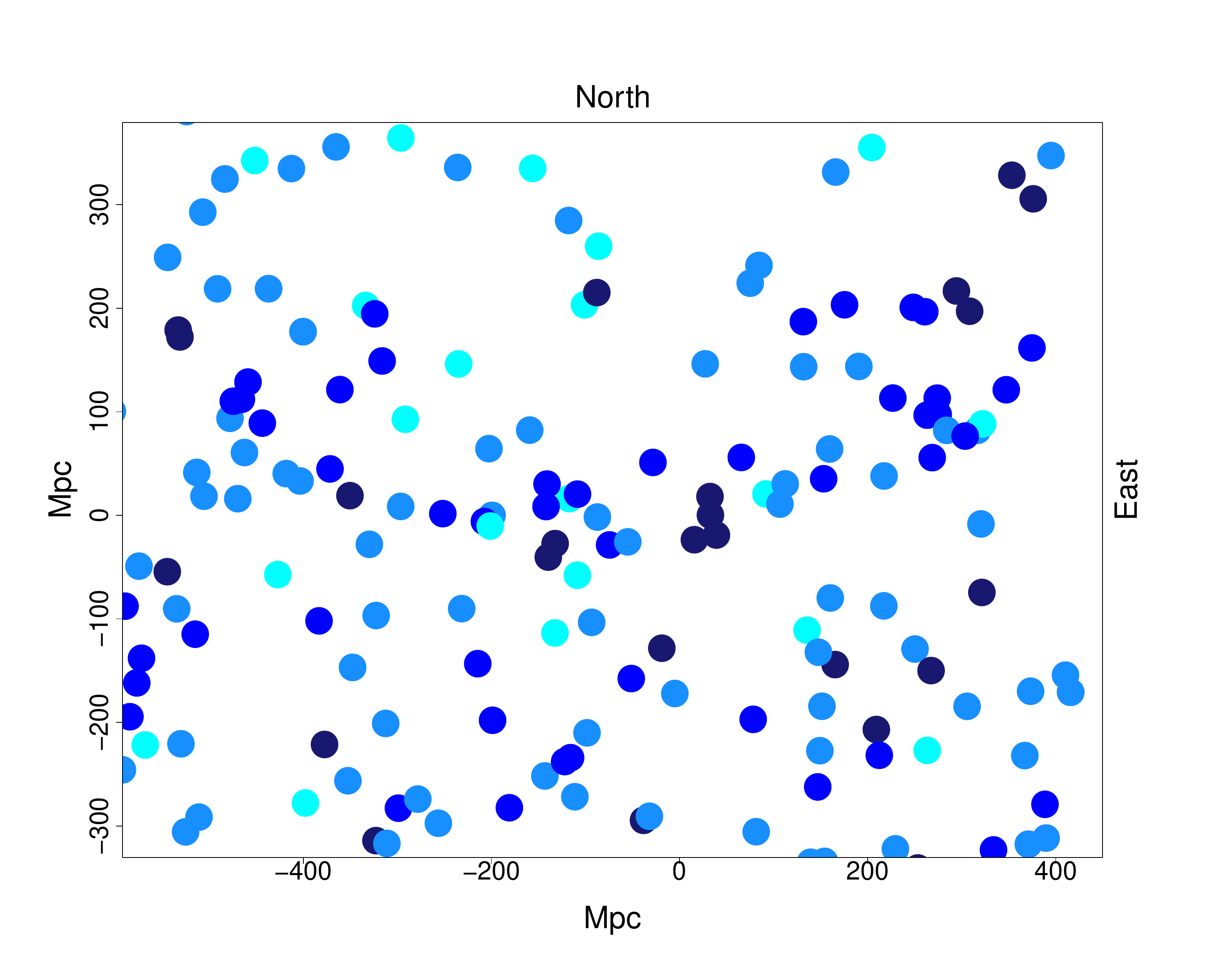}
        \caption {\small The on-sky spatial coordinates of the absorbers in
          the GA and its immediate field, with colour-coding according to the
          four signal-to-noise (S/N) bins of the $\lambda 2796$ Mg~{\sc II}
          line. The four S/N bins are divided: $0, 3, 6, 12, 37$, with the
          lightest shades of blue representing the smallest values. By
          comparing Fig.~\ref{fig:Rplot_abs_EW_col} with the figure here, one
          will notice that the two properties are correlated.}
        \label{fig:Rplot_abs_SN_2796_col}
  \end{subfigure}
    \hfill
   \begin{subfigure}[b]{0.475\textwidth}   
        \centering 
        \includegraphics[width=\textwidth]{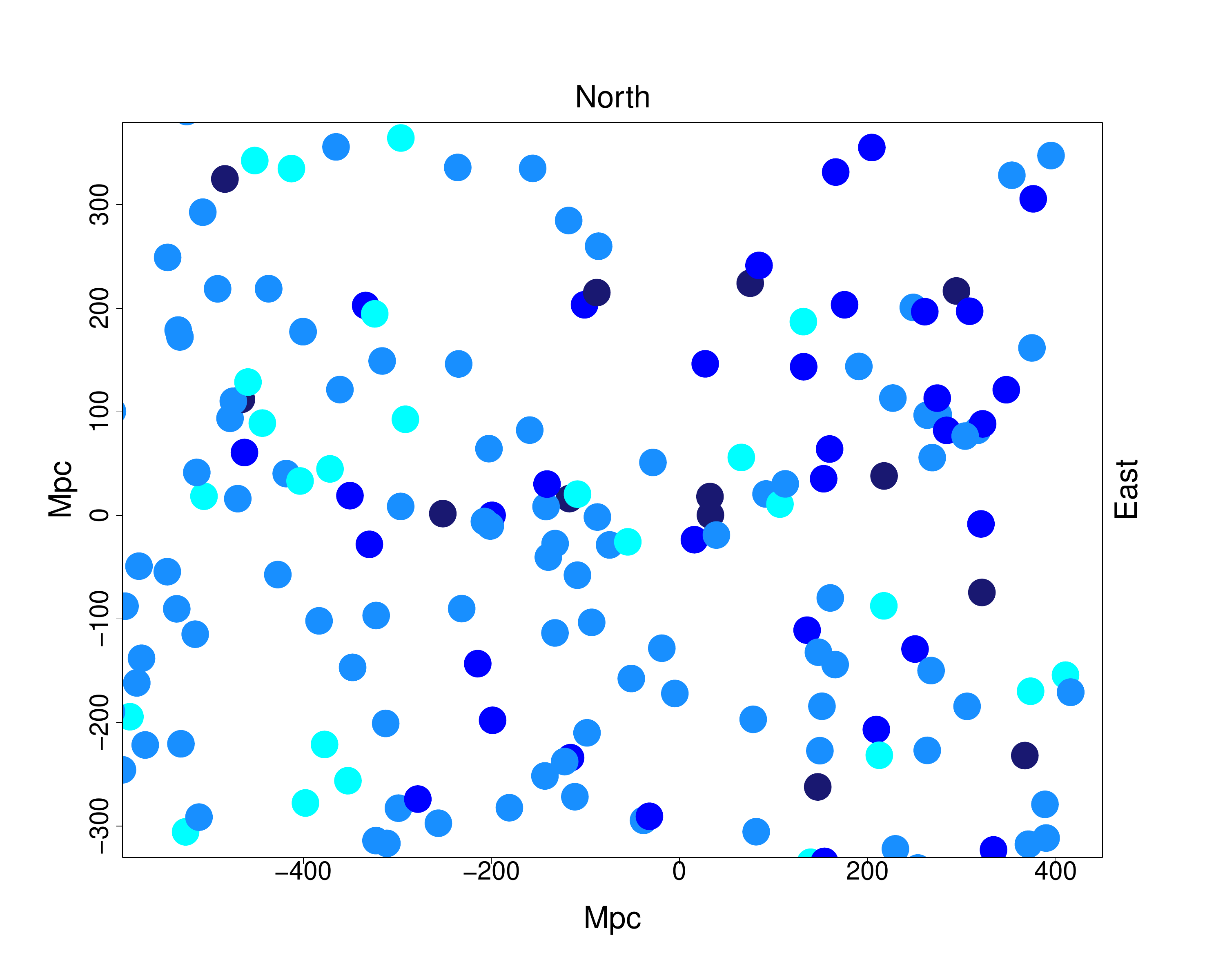}
        \caption {\small The on-sky spatial coordinates of the absorbers in
          the GA and its immediate field, with colour-coding according to the
          four signal-to-noise (S/N) bins of the quasar continuum. The four
          S/N bins are divided: $0, 8, 16, 24, 47$, with the lightest shades
          of blue representing the smallest values. By comparing
          Fig.~\ref{fig:Rplot_abs_imag_col} with the figure here, one will
          notice that the two properties are correlated.}
        \label{fig:Rplot_abs_SN_con_col}
  \end{subfigure}
  \caption{The on-sky spatial coordinates of the absorbers in the GA and its
    immediate field, with colour-coding according to the distribution of:
    equivalent width (EW) of the Mg~{\sc II} $\lambda 2796$ line
    ($W_{r,2796}$); quasar brightness ($i$); signal-to-noise (S/N) of the
    $\lambda 2796$ Mg~{\sc II} line; S/N of the quasar continuum.  The
    lightest shades of blue represent the smallest values (for $i$, this
    means that the lightest shades represent the brightest quasars).}
  \label{fig:Rplot_abs_col}
\end{figure*}

We divide the $W_{r,2796}$ EWs into four bins with boundaries at $0.0, 0.3,
0.6, 1.0, 10.0$\,\AA. (The boundaries were chosen to reflect the above
diversity of what corresponds to `strong' in the literature.) The on-sky
spatial coordinates of the absorbers in the GA and its immediate field are
then plotted, with colour-coding according to the four EW bins --- see
Fig.~\ref{fig:Rplot_abs_EW_col}. The shade of the blue dots in the figure
represents the EW bin, with the lightest shade representing the first bin
$0.0 < W_{r,2796} \leq 0.3$\,\AA, and the darkest shade representing the last
bin $1.0 < W_{r,2796} \leq 10.0$\,\AA.

Similarly, with the same set of four blue shades, we show the S/N of the
$\lambda 2796$ Mg~{\sc II} line, the S/N of the continuum, and the
$i$~magnitude of the probes --- see Figs~\ref{fig:Rplot_abs_SN_2796_col},
\ref{fig:Rplot_abs_SN_con_col}, and \ref{fig:Rplot_abs_imag_col}. The
boundaries of the bins are as follows: (i) S/N of the $\lambda 2796$ Mg~{\sc
  II} line --- $0, 3, 6, 12, 37$; (ii) S/N of the continuum --- $0, 8, 16,
24, 47$; and (iii) $i$ magnitude of the background quasars --- $16.0, 17.8,
18.7, 19.6, 21.0$. The colour-coding again represents the smallest values by
the lightest shade of blue, and the largest values by the darkest
shade. Note, that for $i$, the lightest shade thus represents the brightest
probes.

The EW of the $\lambda 2796$ Mg~{\sc II} line should correlate with the S/N
of the $\lambda 2796$ Mg~{\sc II} line; Figs~\ref{fig:Rplot_abs_EW_col} and
\ref{fig:Rplot_abs_SN_2796_col} show that this is indeed the case. The
brightness ($i$) of the background quasar should correlate with the S/N of
the quasar continuum; Figs~\ref{fig:Rplot_abs_imag_col} and
\ref{fig:Rplot_abs_SN_con_col} show that this is indeed the case. Note that
brighter quasars, having a higher continuum S/N, can detect absorbers to a
lower threshold EW.

An asymmetry is apparent in the distribution of Mg~{\sc II} EWs within the
GA: the EWs tend to be stronger on the LHS (lower RA), and in the centre of
the GA, than on the RHS (higher RA). Conversely, there is a tendency for the
RHS to have brighter probes and higher continuum S/N, so, on the RHS, the
threshold EW for detection should tend to be lower and certainly able to
detect stronger absorbers, should they be present. Thus, the collected
observations are consistent with the reality of the observed asymmetry of
stronger absorbers on the LHS.

As discussed earlier, in Section \ref{MgII_approach}, there have been many
attempts to understand the relationship between EW and galaxy properties
(morphology, luminosity, impact parameter, galaxy inclination etc.), but so
far without a clear understanding of the connections between
them. Conceivably, the asymmetry in the EW distribution could arise from the
details of the geometry of the GA and the orientations of the galaxies within
it. Future sky surveys and targeted observations seem likely to be necessary
for progress on these details.

We note that there appears to be a preference for the strongest
($W_{r,2796}$) Mg~{\sc II} absorbers in the GA to clump together into groups
of a few. See the dark blue points in Fig.~\ref{fig:Rplot_abs_EW_col}, and
note in particular those on the LHS of the GA (lower RA), the centre of the
GA, and the group just above the tip of the RHS (higher RA) of the GA. As the
GA is denser than the rest of the field, we can speculate that the occurrence
of the strongest EWs in proximity is not accidental but is connected with the
origin and environment of the GA.

Recall that the SZ cluster B18, $z = 0.763$ \citep{Burenin2018}, is at the
centre of the GA. (It is what led to the discovery of the GA.) At the centre
of the GA is a small, circular `hole', and surrounding this hole is a group
of the stronger absorbers. SZ clusters create a highly-ionised environment,
but Mg~{\sc II} absorption occurs in low-ionised regions. Possibly, a region
of high ionisation can account for the hole, but an origin, in environment,
of the enveloping group of stronger absorbers is not then obvious.

The investigation of small ($\Delta z = 0.030$), overlapping (by 50 per cent)
redshift slices reveals a noticeable difference between the left and right
hand sides of the GA. For example, the LHS (lower RA) of the GA appears
concentrated in the small redshift slice located farthest away ($z = 0.832
\pm 0.030$), whereas the RHS (higher RA) of the GA appears spread diffusely
through the larger redshift slice.  Interestingly, the LHS of the GA has both
a narrower redshift distribution and a preference for stronger Mg~{\sc II}
EWs.

Finally, the investigation of the redshift distribution suggests that, if the
GA is represented as a segment of a cylindrical shell, then the LHS would be
tilted away along the line of sight. That is, if the GA can indeed be
represented as a segment of a cylindrical shell, then it is not precisely
orthogonal to the line-of-sight but is rotated with respect to a north-south
axis.

\subsection{Connectivity and statistical properties}

The GA was discovered visually, from a Mg~{\sc II} density image
(e.g.\ Fig.~\ref{fig:The_GA_MgII_image}). Albeit after the event, we now
discuss its connectivity and statistical properties. The Mg~{\sc II}
absorbers can, of course, be found only where there are background quasars to
act as probes, and those probes may themselves be subject to spatial
variations arising from large-scale structure and, in particular, from
artefacts in the surveys.

We apply three different statistical methods for assessing the GA, as
follows.

(i) SLHC / CHMS --- see \citet{Clowes2012}. This method depends first on
constructing the 3D minimal spanning tree (MST), and then separating it at
some specified linkage scale. At this stage it is equivalent to
single-linkage hierarchical clustering (SLHC). The statistical significance
of a candidate structure is then assessed using its volume obtained as the
volume of the `convex hull of member spheres' (CHMS). Note the important
feature that this method assesses the significance of {\it individual}
candidate structures.

(ii) The Cuzick-Edwards (CE) test --- see \citet{CuzickEdwards1990}. It is a
2D `case-control' method that is designed to correct the incidence of cases
for spatial variations in the controls (the underlying population). It
depends on the number of cases that occur within the $k$ nearest
neighbours. The CE test can detect the presence of clustering in the field,
while correcting for variations in the background, and can assess its
statistical significance. It cannot, however, assess the physical scale of
the clustering.

(iii) 2D Power Spectrum Analysis (2D PSA) --- see \citet{Webster1976a}. It is
a powerful Fourier method for detecting clustering in the field. It can be
effective even for detecting weak clustering. The 2D PSA can detect the scale
of clustering and assess the statistical significance of the clustering at
that scale.

Each of these tests has distinctive attributes, and the reader should judge
the evidence provided by the ensemble. Only the SLHC / CHMS method assesses
the significance of individual candidate structures, whereas the CE test and
the 2D PSA address clustering in the field. We shall describe below the
`polygon approach', in which we assess the contribution that the GA makes to
the results from the CE test and the 2D PSA for the field. Only the CE test
can correct for spatial variations in the underlying population. However, we
shall describe below, again using the polygon approach, that the 2D PSA has
more power to discriminate than the CE test.

Finally, we emphasise again, that given the nature of the discovery the
statistical analysis is necessarily performed {\it post-hoc}.  The reader
will find that we have used techniques to compare the field containing the GA
with other, unrelated fields (within the same Mg~{\sc II} dataset). This of
course has its limitations due to the non-uniformity of the background
quasars (probes) and potential survey artefacts. We have also compared with
randomised simulations, in which we attempt to preserve these subtleties of
the Mg~{\sc II} data.

\subsubsection{SLHC / CHMS (Minimal Spanning Tree)}
\label{subsect:MST}

The minimal spanning tree (MST) is a widely-used algorithm for assessing
large-scale structure in astronomy and cosmology. When the MST is separated
at some specified linkage scale it is equivalent to the algorithm for
single-linkage hierarchical clustering (SLHC). An approach to assessing the
statistical significance of the agglomerations found in this way was
introduced by \citet{Clowes2012}: the Convex Hull of Member Spheres (CHMS)
method. It was further used by \citet{Clowes2013} in the analysis of the
Huge-LQG, the Huge Large Quasar Group that they discovered.

Here, we apply the sequence of SLHC and CHMS to the Mg~{\sc II} absorbers in
the GA field. By specifying a linkage scale and a minimum membership, the
SLHC identifies the 3D agglomerations or groups within the coordinates of the
absorbers. Within each identified group the CHMS constructs a sphere around
each member point with a radius of half the mean linkage separation for that
group. A volume for the group is then computed as the volume of the convex
hull of its member spheres (and note that the convex hull is a unique
construction). An expected density of absorbers is determined from a control
field and the observed redshift interval of a group. (Here, the control field
is specified as the same field that is being assessed.) The observed number
of member points within a group are then scattered randomly within a cube at
the expected density, and their CHMS volume is calculated; this is done 1000
times. The significance of the group is calculated by the rate of occurrence
of randomly-generated CHMS volumes that are smaller than the observed
volume. See \citet{Clowes2012} for full details of the CHMS method.

In principle, this SLHC / MST approach should be applied only to surveys that
have no intrinsic spatial variations. The background quasars --- the probes
of the Mg~{\sc II} absorbers --- are drawn from a merger of the SDSS DR7QSO
and DR12Q databases. While a reasonably spatially-uniform subset can be
extracted from DR7QSO, DR12Q is much more strongly affected by spatial
artefacts arising from deeper areas. Thus the distribution of the background
quasars can conceivably affect the distribution of the Mg~{\sc II} absorbers
in some, possibly complicated, way. However, if the distribution of the
background quasars appears to be reasonably homogeneous in the area of
interest, then we can assume that the distribution of Mg~{\sc II} absorbers
is predominantly a product of the LSS and not the availability of background
quasars. Of course, the distribution of background quasars can still have
some effects --- such as occasional gaps in connectivity --- on the Mg~{\sc
  II} absorbers even in such reasonably homogeneous regions.

\begin{figure}
    \centering
    \includegraphics[scale=0.18]{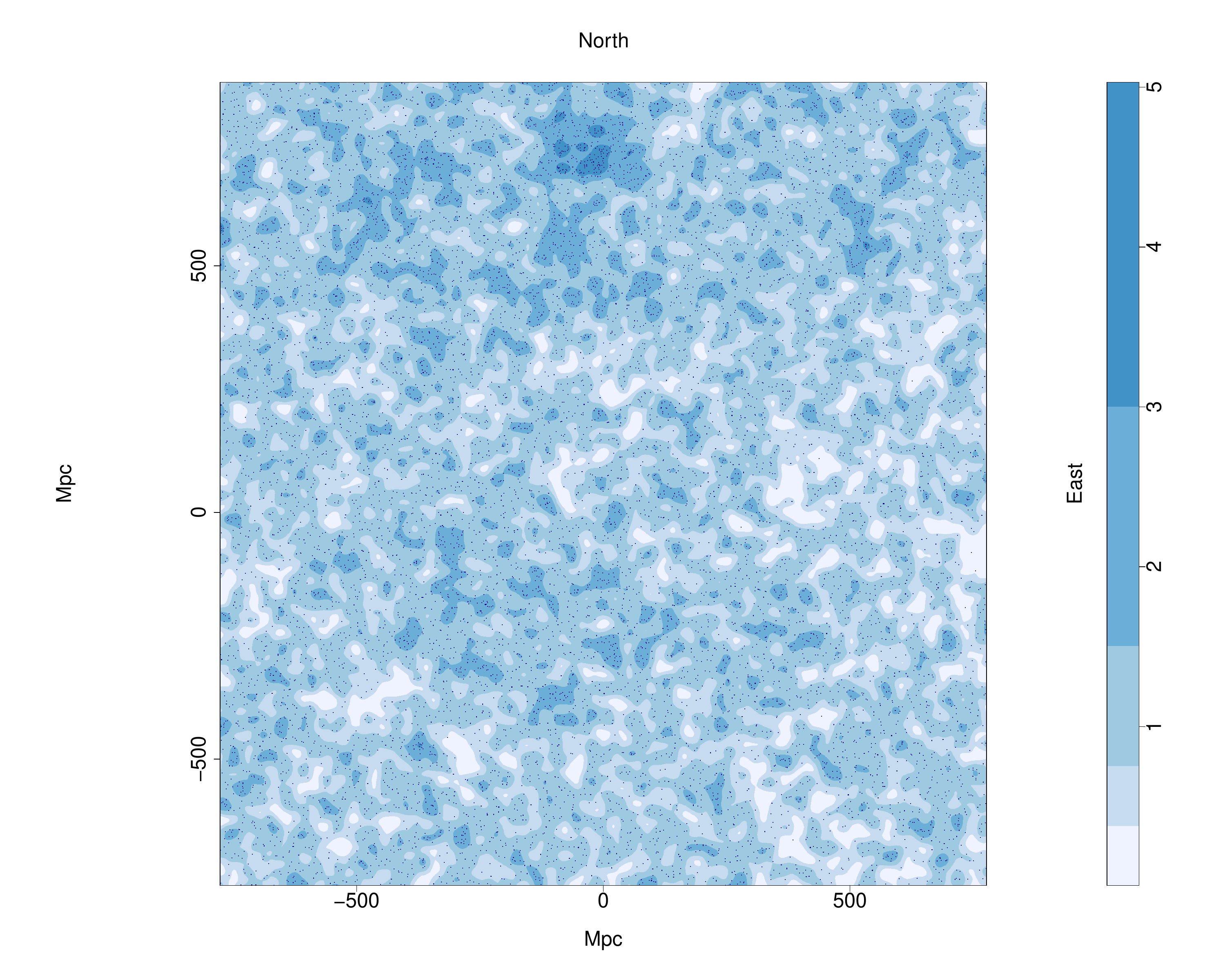}
    \caption{The tangent-plane distribution of background probes (quasars) 
      in the GA field with redshifts $z > 0.862$, represented by the
      dark blue points. The blue contours, increasing by a factor of two, 
      represent the density distribution
      of the quasars using a Gaussian kernel with a smoothing scale of
      $\sigma = 11$~Mpc. The axes are labelled in Mpc, scaled to the present
      epoch. East is towards the right and north is towards the top. 
      It is clear that there are denser
      areas, less dense areas, and even empty patches, across the whole
      image, indicating the spatial non-uniformity of the background probes.
      There is a particularly dense band in approximately the northern third.
      However, there are evidently no artefacts that correspond to the
      dimensions and orientation of the GA.}
    \label{fig:qso_field_x3}
\end{figure}

Fig.~\ref{fig:qso_field_x3} shows the kernel-smoothed distribution
($\sigma = 11$~Mpc, present epoch) of the background probes (quasars) in the
area of the GA for $z > 0.862$. It is clear that there are denser areas,
less dense areas, and even empty patches, across the whole image, indicating
the spatial non-uniformity of the background probes. There is a particularly
dense band in approximately the northern third, which arises from a deeper
area of the DR12Q survey. However, there are evidently no artefacts that
correspond to the dimensions and orientation of the GA.

We are now taking the GA to be predominantly concentrated in the redshift
interval $0.802 \pm 0.060$, so $0.742 \rightarrow 0.862$. Its (present-epoch)
depth is then $\sim 340$~Mpc. 

This redshift interval appears to be the optimum, following a heuristic
process of stepping through a range of redshift intervals and determining the
membership and significance of the GA through the SLHC / CHMS
method. Redshift intervals of thickness $\Delta z = \pm 0.050$ between $0.760
< z < 0.810$ were tested, using various linkage scales, and a clear peak
signal for the GA was seen around the redshift $z = 0.810$ using a linkage
scale of $95$~Mpc. (See Section \ref{subsect:linkage_scale} for the details
of choosing the optimum linkage scale.) Further, finer-scale, testing
revealed the greatest number of connected GA members to be more precisely
located at $z = 0.802$, again for the linkage scale of $95$~Mpc. The
significance for $z = 0.802 \pm 0.050$ is $4.15\sigma$. Widening the redshift
interval to $z = 0.802 \pm 0.060$ gives a slightly higher significance of
$4.30\sigma$, and the number of member Mg~{\sc II} absorbers is then
increased from 42 to 44.

A second, smaller, agglomeration made up of 10 and 11 absorber members at
both $z = 0.802 \pm 0.050$ and $z = 0.802 \pm 0.060$ respectively, although
not formally significant ($1.75\sigma$ and $2.04\sigma$ respectively), is
clearly also part of what we identified visually as the Giant
Arc. Conceivably, just one further background probe would be sufficient to
yield one further absorber that would then connect both agglomerations as one
significant unit. We have emphasised the limitations of the SLHC / CHMS
method for this dataset, and we might here be seeing their consequences.

As noted above, the estimation of the CHMS significance requires a control
field from which the expected average density is calculated. For the CHMS
significances given above, we used the same field as that containing the GA.
This was a deliberate choice, given the spatial variations of the wider
survey. Clearly, the GA must then represent a small fraction of the total
area and number of absorbers ($\sim$ 7 per cent of the absorbers
are from the GA).

\begin{figure}
    \centering
    \includegraphics[scale=0.18]{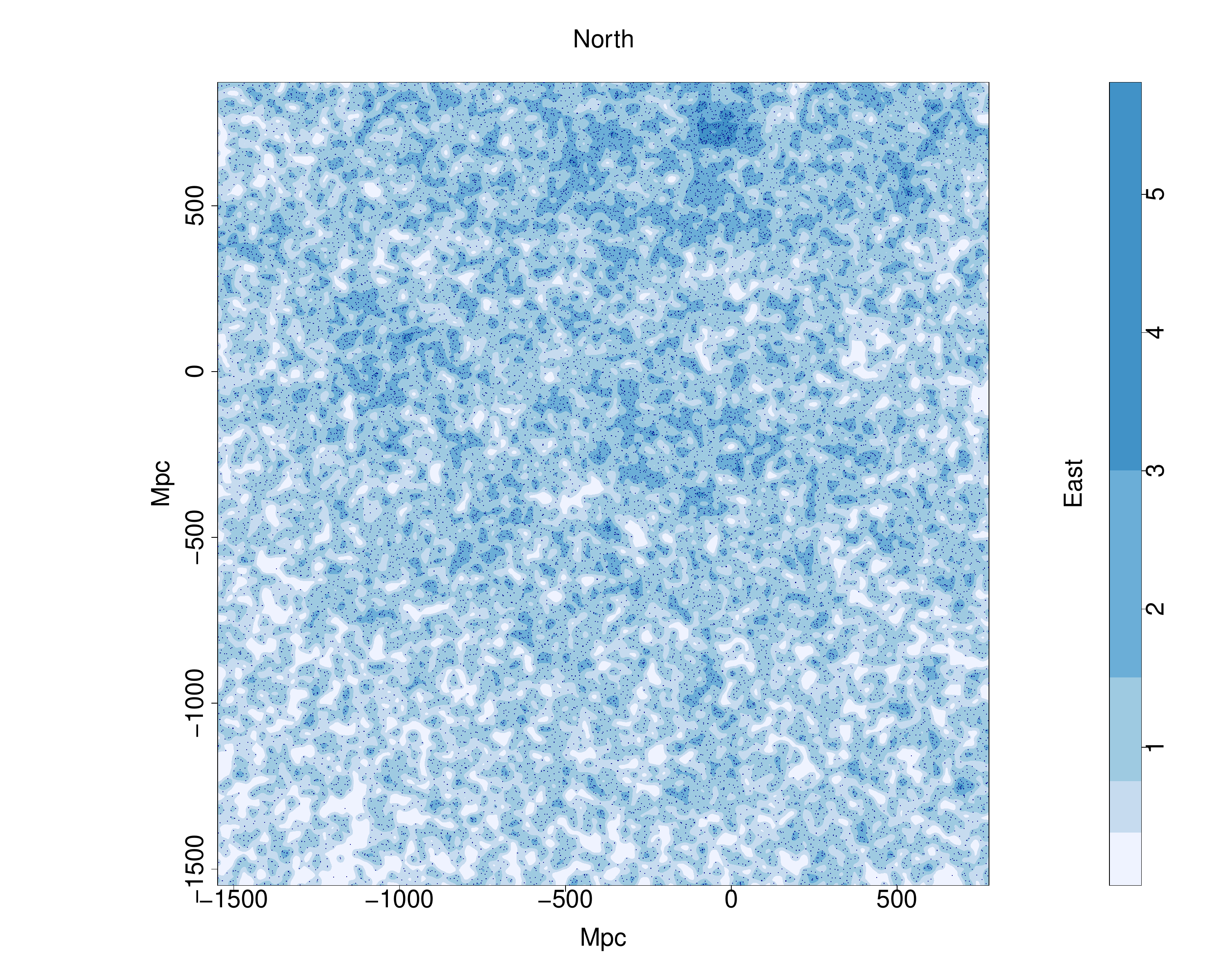}
    \caption{The tangent-plane distribution of background probes (quasars),
      in the same region of sky as the GA for redshifts $z > 0.862$,
      represented by the dark blue points. The field-of-view has here been
      increased by extending the western and southern boundaries. The blue
      contours, increasing by a factor of two, represent the density
      distribution of the quasars using a Gaussian kernel with a smoothing
      scale of $\sigma = 11$~Mpc. The axes are labelled in Mpc, scaled to the
      present epoch. East is towards the right and north is towards the top.}
    \label{fig:qso_bigger_FOV}
\end{figure}

Even so, the distribution of background probes (quasars) in the field of the
GA is not uniform --- notably the denser band in the northern third.  This
non-uniformity could affect the CHMS calculations of significance, either by
overestimating or underestimating, depending on whether the probes are
generally under-populated or over-populated in the control field. A second
estimate of the significance can be calculated from the CHMS method by
increasing the field-of-view (FOV) containing the GA, and using it as a new
control field. Fig.~\ref{fig:qso_bigger_FOV} shows the background probes in
the field containing the GA with the western and southern boundaries
extended. Note that the eastern and northern boundaries were not extended
because of proximity to the edge of the survey area.

\begin{figure}
    \centering
    \includegraphics[scale=0.4]{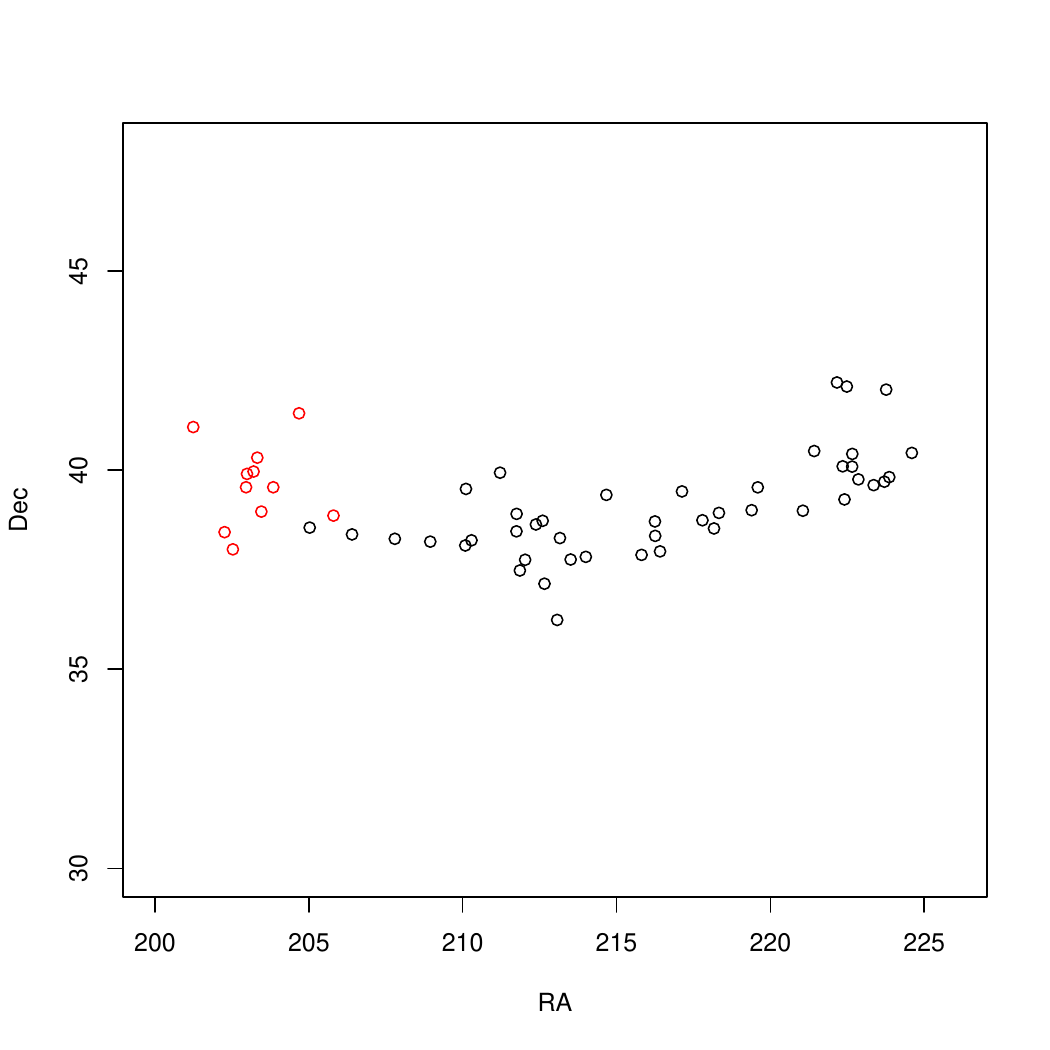}
    \caption{The GA connected via the SLHC algorithm. There are 44 black
      points which indicate the principal agglomeration of the Giant Arc ---
      the largest, and most significant agglomeration in the field ---, which
      comprises the majority of the GA. There are 11 red points which
      indicate the secondary agglomeration of the GA, which, although by
      itself not statistically significant, is clearly part of the GA. The
      axes are labelled RA and Dec where east is towards the right and north
      is towards the top.}
    \label{fig:CHMS_GA_bigger_FOV}
\end{figure}

Using the {\it larger} FOV in the redshift interval $z = 0.802 \pm 0.060$,
the CHMS method calculates a significance of $4.53\sigma$ for the principal
agglomeration of the GA. As noted previously, the GA is split into two
agglomerations by the SLHC algorithm, shown in
Fig.~\ref{fig:CHMS_GA_bigger_FOV}. For this entire, larger, FOV,
there are 35 agglomerations in total, with the principal agglomeration of the
GA being the largest and most significant, and the only agglomeration with a
significance $> 3.5\sigma$. Fig.~\ref{fig:CHMS_GA_bigger_FOV} shows the GA as
located by the SLHC / CHMS method, with the principal agglomeration
represented by the black points; the red points indicate the smaller,
separate agglomeration of much lower significance, but visually it can
clearly be seen as part of the GA.

The SLHC / CHMS algorithm is next applied to three lower redshift slices ---
$z = 0.682 \pm 0.060$, $z = 0.562 \pm 0.060$, and $z = 0.442 \pm 0.060$ ---
in the same (larger) FOV as the GA (see Fig. \ref{fig:qso_bigger_FOV}). Using
lower redshift slices and the same FOV means that we can conveniently compare
the Mg~{\sc II} absorbers arising from the same probes as those in the GA
field by restricting the probes to $z > 0.862$. We can also apply the SLHC /
CHMS method to the same three lower redshift slices {\em without} this
redshift restriction on the probes. Recall, however, that the probes may show
density artefacts with a large FOV.  As mentioned previously, the SLHC / CHMS
method can be problematic for the Mg~{\sc II} analysis because of these
artefacts. Therefore, superficially significant structures that correspond to
particularly dense patches of probes are likely to be discarded. The results
are summarised in Table~\ref{tab:zslices_list}.

\begin {table*}
\flushleft
\caption{Results from applying the SLHC / CHMS algorithm to three lower
redshift slices in the same (larger) FOV as the GA (see
Fig. \ref{fig:qso_bigger_FOV}). The columns are: the redshift interval of
the Mg~{\sc II} absorbers; the redshift of the background probes (quasars);
the number of structures found by the SLHC algorithm; the number of
structures exceeding $3 \sigma$ significance; the number of structures
exceeding $4 \sigma$ significance; the number of structures exceeding $3
\sigma$ significance after removing those that reside in artefacts of the
probes. Note that the three lower redshift slices have two sets of probes
with different redshifts: one for the probes corresponding to those in the
GA field, and one for the probes corresponding to the redshift slice.
}
\small \renewcommand \arraystretch {0.8}
\newdimen\padwidth
\setbox0=\hbox{\rm0}
\padwidth=0.3\wd0
\catcode`|=\active
\def|{\kern\padwidth}
\newdimen\digitwidth
\setbox0=\hbox{\rm0}
\digitwidth=0.7\wd0
\catcode`!=\active
\def!{\kern1.3\digitwidth}
\begin {tabular} {llllll}
\\
Redshift slice      & Probes        &  N total & $N \ge 3.0\sigma$ & $N \ge 4.0\sigma$ & N not rejected \\
                    &               &          &                   &                   & as artefacts   \\
\\
$z=0.802 \pm 0.060$ & $z > 0.862$   & 35       & 4                 & 1 (GA)            & 2 (incl. GA)   \\
$z=0.682 \pm 0.060$ & $z > 0.862$   & 31       & 3                 & 1                 & 3              \\
                    & $z > 0.742$   & 34       & 6                 & 2                 & 4              \\
$z=0.562 \pm 0.060$ & $z > 0.862$   & 17       & 6                 & 3                 & 2              \\
                    & $z > 0.622$   & 19       & 7                 & 4                 & 3              \\
$z=0.442 \pm 0.060$ & $z > 0.862$   & 9        & 1                 & 0                 & 0              \\
                    & $z > 0.502$   & 12       & 2                 & 0                 & 0              \\
\\
\end {tabular}
\\
Notes. \\
(1) For each redshift slice the overall density of absorbers in the field
varies, so we modify the linkage scale according to the relation $s = (\rho_0
/ \rho)^{1/3} \times 95\,\mathrm{Mpc}$, where $s$ is the linkage scale,
$\rho_0$ is the density for the GA field and $\rho$ is the density for the
new field concerned. \\
(2) The number of structures reduces in each successively lower redshift
slice, because the number of Mg~{\sc II} absorbers reduces as the
lower-wavelength limit of detectability is approached.
\label{tab:zslices_list}
\end {table*}

As can be seen in Table~\ref{tab:zslices_list} there are significant ($\sim
3\sigma$) structures. At a more cautious limit of $4\sigma$, however, there
were only two candidate structures which did not reside in an artefact (in
redshift slices $z=0.682$ and $z=0.562$). We shall in due course investigate
them further, starting with optimisation of the redshift intervals.

Finally, we introduce a random-simulation aspect to the SLHC / CHMS
analysis. We have carried out 1000 random simulations as follows. (i) We
consider the large, extended area that corresponds to
(Fig. \ref{fig:qso_bigger_FOV}). (ii) We consider only the probes at higher
redshift than the redshift slice of the GA --- that is, we continue (as with
the slices of redshift lower than that of the GA) to use only the probes
appropriate to the GA, so that density artefacts in the probes remain
identical. (iii) We reassign at random MgII absorbers of any redshift to the
probes, while not splitting occurrences of multiple absorbers per line of
sight. (Note that splitting absorbers would have the undesirable effect of
changing the total number of probes with absorbers.) (iv) We then analyse the
random-simulated data as for the actual GA slice, selecting absorber
redshifts for the redshift slice.

Within the simulations, we looked for ``structures'' that had properties
comparable to, or more extreme than, the observed properties of the GA
(precisely, of GA-main --- see below). The properties considered were the set
of: number of members; SLHC / CHMS significance; and overdensity. In all
cases (roughly one occurrence per two simulations), we found that these
``comparable structures'' were in the regions of the visually-obvious density
artefacts, and never in the region occupied by the real GA. The occurrence of
the ``comparable structures'' in the density artefacts is as expected: for
those artefacts, the linkage scale and the control density would clearly not
be appropriate. We can infer that the probability of the real GA (precisely,
GA-main) occurring as a random event is $< 0.001$.

The SLHC algorithm easily identifies the GA, with 44 connected Mg~{\sc II}
absorbers, and the CHMS method estimates a significance of $\sim 4.5\sigma$
using the central redshift $z = 0.802$.  In every redshift interval
investigated, the GA appears in two parts: (i) the principal agglomeration,
which is large in both physical size and membership, and statistically very
significant; and (ii) the secondary agglomeration, 
small in size and membership, and by itself statistically not significant. 
As mentioned earlier, the Mg~{\sc II} absorbers depend on the availability of 
background probes (quasars), and without those, Mg~{\sc II} would not be 
detected. Thus an artefact in the distribution of probes --- i.e.\ a gap, 
perhaps of just one missing probe --- could lead to an artefact of apparent
splitting into two agglomerations. 

We have seen previously, in Section \ref{sec:observational_properties_GA},
that there is a noticeable difference between the LHS and RHS of the GA with
regards to redshift distribution. Investigating small ($\Delta z = 0.030$),
overlapping (by 50 per cent) redshift slices has highlighted the
sub-structure of the GA along the redshift axis. We find that the larger
agglomeration of the GA is distributed more evenly and widely along the
redshift axis, while the smaller agglomeration is concentrated in a narrower
redshift slice. It becomes clear from the central redshift slice and below
($z <0 .802$) that there are no Mg~{\sc II} absorbers available that can
connect the small agglomeration to the large agglomeration.

More data, such as the new SDSS DR16Q quasar database \citep{Lyke2020}, could
provide additional information to investigate the GA further. This includes,
conceivably, the possibility of connecting the small agglomeration to the
large agglomeration of the GA.  However, this would require construction of a
new Mg~{\sc II} catalogue corresponding to the DR16Q quasars. In the future,
we plan to create our own Mg~{\sc II} catalogues from previous and future
quasar data releases.

\subsubsection{Selecting a Linkage Scale}
\label{subsect:linkage_scale}

The linkage scale that is set in the SLHC / CHMS method determines both the
number of agglomerations and their memberships. It was set at $95$~Mpc for
the GA. This setting was partly guided by the linkage scale that was known
to be effective for the CCLQG, and which, when used subsequently, led to the
discovery of the Huge-LQG and many other LQGs (see \citet{Clowes2012} and
\citet{Clowes2013}). Clearly, the linkage scale must be adjusted for field
density; in this case, starting from the linkage scale that was effective for
LQGs (i.e.\ for LSS in quasars) we calculate a linkage scale of $85$~Mpc for
Mg~{\sc II} absorbers in the GA field. From here we followed the heuristic
process described in Section \ref{subsect:MST} to identify an optimum linkage
scale of $95$~Mpc for the Mg~{\sc II} absorbers.

One must remember that Mg~{\sc II} absorbers are distinctly different from
quasars and therefore cannot be treated in quite the same way. For example,
the linkage scale that works for quasars will not necessarily work for the
Mg~{\sc II} absorbers since the latter is a case of inhomogeneities (the
absorbers) superimposed on inhomogeneities (the quasars and survey
artefacts). Future work will address the development of a clustering analysis
that is specifically addressed to the requirements of Mg~{\sc II} absorbers.

It is in the nature of discoveries that there will be a {\it post-hoc} aspect
to the analysis. What turned out to be effective for the CCLQG led to the
discovery of the Huge-LQG: an initial discovery, followed by a heuristic
process, followed by an entirely objective {\it a-priori} new discovery.  It
is in this spirit that we present the discovery of the GA: one for which the
techniques and parameters used to assess and characterise it can subsquently
be applied to the whole Mg~{\sc II} database.

\begin{figure}
    \centering
    \includegraphics[scale=0.15]{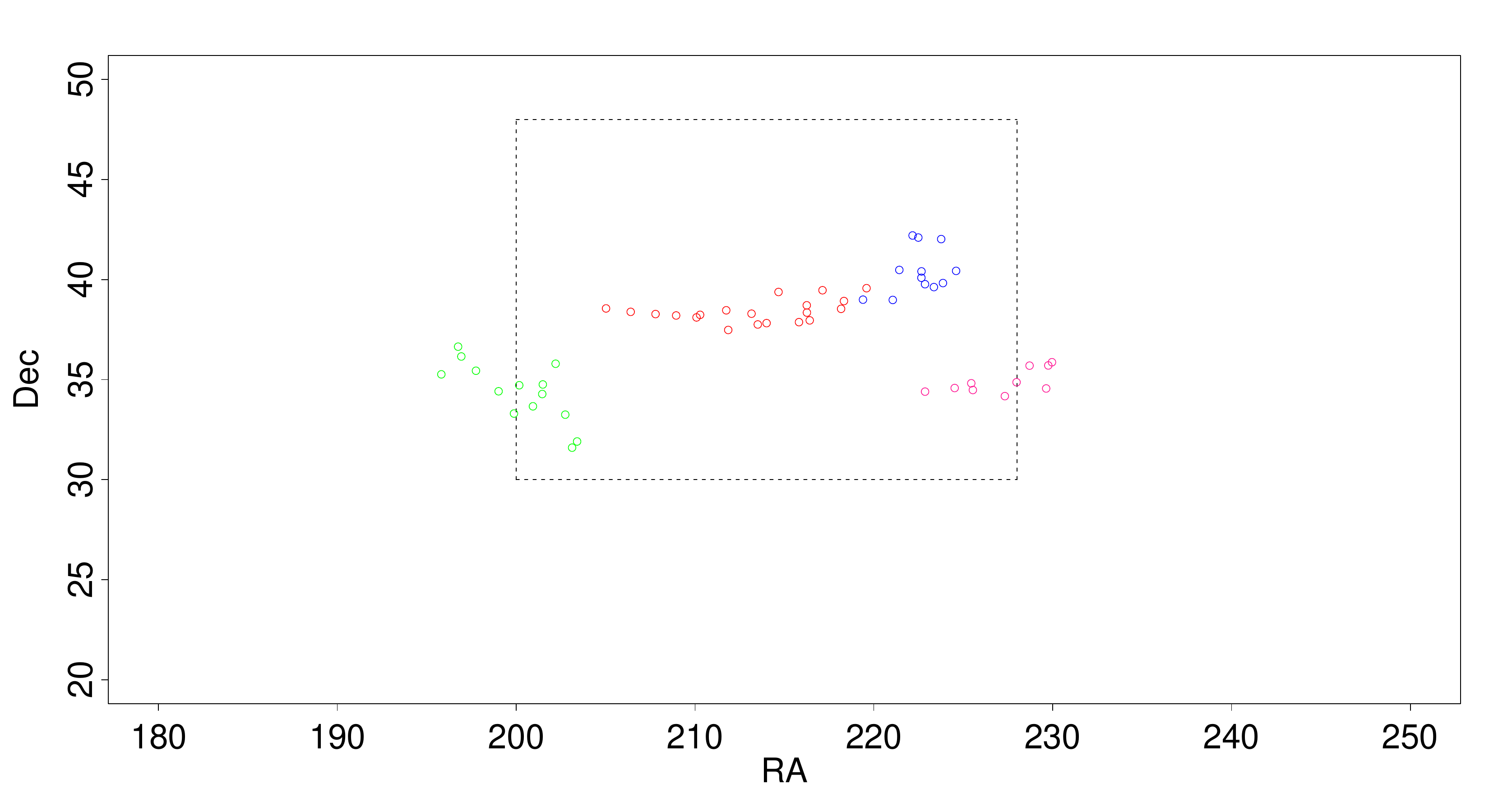}
    \caption{The four structures found via the SLHC / CHMS method using a
      reduced linkage scale of $85$~Mpc. The black dashed rectangle
      corresponds to Fig. \ref{fig:CHMS_GA_bigger_FOV} for comparison. The GA
      is still mostly detected in what appears to be two parts. The largest
      agglomeration, containing $20$ Mg~{\sc II} absorber members represented
      by the red points in the figure, makes up the majority of what we
      visually identify as the GA. The second largest agglomeration,
      containing $14$ Mg~{\sc II} absorber members represented by the blue
      points in the figure, is also part of what we visually identified as
      the GA and can be seen as an extension of the red points.  The axes are
      labelled RA and Dec where east is towards the right and north is
      towards the top.}
    \label{fig:CHMS_85_GA_bigger_FOV}
\end{figure}

For completeness, we can briefly mention what results from instead setting
the linkage scale to $85$~Mpc, $90$~Mpc, $100$~Mpc, and $105$~Mpc for the
adopted redshift slice. In five runs, using linkage scales of $85$~Mpc,
$90$~Mpc, $95$~Mpc, $100$~Mpc, and $105$~Mpc, there are totals of $4$, $25$,
$35$, $43$, and $3$ agglomerations found respectively, with the GA always
being the most significant in all except the $105$~Mpc run. The middle three
runs split the GA into two parts --- one large and significant, and one
smaller and less significant. For the following comments, we concentrate only
on the large, significant part of the GA located in the middle three runs
($90$, $95$ and $100$~Mpc), which makes up the majority of what we visually
identified as the GA. (1) It has a significance greater than $3.8\sigma$ in
all of the three runs. (2) It is the only agglomeration that has a
significance greater than $3.5\sigma$, with only two or three agglomerations
above $3\sigma$ (all others being below a $3\sigma$ threshold). (3) In both
the $90$~Mpc and $95$~Mpc runs, it is the largest agglomeration by
membership, and is the second largest by membership in the $100$~Mpc run.
Lastly, we mention what arises from setting the linkage scale to $85$~Mpc and
$105$~Mpc. Using these linkage scales one can see that the SLHC / MST method
has reached its maximum and minimum limit with the linkage scales, as there
are only $4$ and $3$ structures found, respectively. The corresponding
memberships are $20$ and $133$ with significances of $3.6\sigma$ and
$0.8\sigma$, respectively. As is demonstrated, going any lower or higher with
the linkage scale would yield nothing of consequence: either no structures,
or one large structure containing almost everything. It is worth noting that,
although the $85$~Mpc linkage scale is the minimum at which structures can be
found in the Mg~{\sc II} data at this redshift / density, the GA is still
mostly detected, still the largest agglomeration in the field, and the only
structure detected with a significance over $3\sigma$ --- see
Fig. \ref{fig:CHMS_85_GA_bigger_FOV}

Note that for future work, involving the remainder of the Mg~{\sc II}
database, we can adopt this scale of $95$~Mpc as a standard, with scaling
according to the density of absorbers (i.e., $s = (\rho_0 / \rho)^{1/3}s_0$)
in the volumes of interest.

\subsubsection{Cuzick-Edwards test}

We mentioned above that, strictly, the SLHC / MST approach should be applied
only to surveys that have no intrinsic spatial variations. However, a
statistical test for (two-dimensional) clustering exists that is designed to
manage spatial variations in the source data: the Cuzick-Edwards test
\citep{CuzickEdwards1990}. We apply it here.

The Cuzick-Edwards test (hereafter CE test) has been used mainly in medical
research, such as the clustering patterns of diseases within unevenly
populated geographical regions. (The essential character of our problem is
the same.) It adopts a `case-control' approach to a $k$ nearest-neighbour
(NN) analysis. Several papers have compared the properties of the CE test
amongst various spatial clustering analyses and assert that the CE test is
powerful and sensitive in estimating clustering significance within a point
dataset --- see, for example, \citet{Song2003}, \citet{Meliker2009} and
\citet{Hinrichsen2009}. In \citet{Song2003} the authors note that the CE test
is used more appropriately if the level of clustering is known beforehand.

Inevitably, for our problem, the statistical properties of the GA are tested
after the event of discovery (i.e.\ the level of clustering is known).

We used the CE test that is coded in the application {\it qnn.test} in the
{\sc R} package {\sc smacpod} \citep{French2020} --- {\it Statistical Methods
  for the Analysis of Case-Control Point Data}. The probes (i.e.\ the
background quasars) are labelled as `controls' and the Mg~{\sc II}
absorbers in the redshift interval are labelled as `cases'. The {\it
  qnn.test} then uses a NN algorithm to find the $q$ (or $k$) NNs of any case
to another case.

The test statistic is then calculated as

$$ T_k = \sum_{i=1}^n \delta_id_i^k $$

\noindent
where
\newline
$\delta_i = 1$ if the data point is a case or $0$ if it is a control;
\newline
$d_i^k = 1$ if the NN is a case and $0$ if it is a control.

The $p$-value from {\it qnn.test} is calculated from simulations under the
random-labelling hypothesis \citep{French2020} for $\mathrm{nsim} = 2000$
simulations.

The choice of maximum $q$ ($k$) value that is adopted for the test will
depend on the control-case ratio, as can be seen from the test statistic
calculation. There are $\sim 20$ times as many probes (controls) as Mg~{\sc
  II} absorbers (cases) in the redshift interval of the GA.
\citet{CuzickEdwards1990} examine the power of the CE test with varying
control-case ratios and find that a control-case ratio of between $4$ and $6$
is optimum (see their Fig.~5).

Therefore, we choose to use a control-case ratio of $\sim$~5:1.  To achieve
this we randomly select 25 per cent of the probes, for each of 100 runs of
{\it qnn.test}. (Randomly-selected controls that duplicate the coordinates of
the cases in a given run are removed.) The 100 runs also allow us
to assess how robust are the estimates of significance for the Mg~{\sc II}
absorbers.

We use a set of $q$ ($k$) values: 1, 2, 4, 8, 12, 16, 20, 24, 28, 32, 36, 40, 
44, 48, 52, 56, 60, and 64.
We start by applying the {\it qnn.test} to the basic GA field.

\begin{figure*}
      \centering
      \begin{subfigure}[b]{0.5\textwidth}
        \centering
        \includegraphics[width=\textwidth]{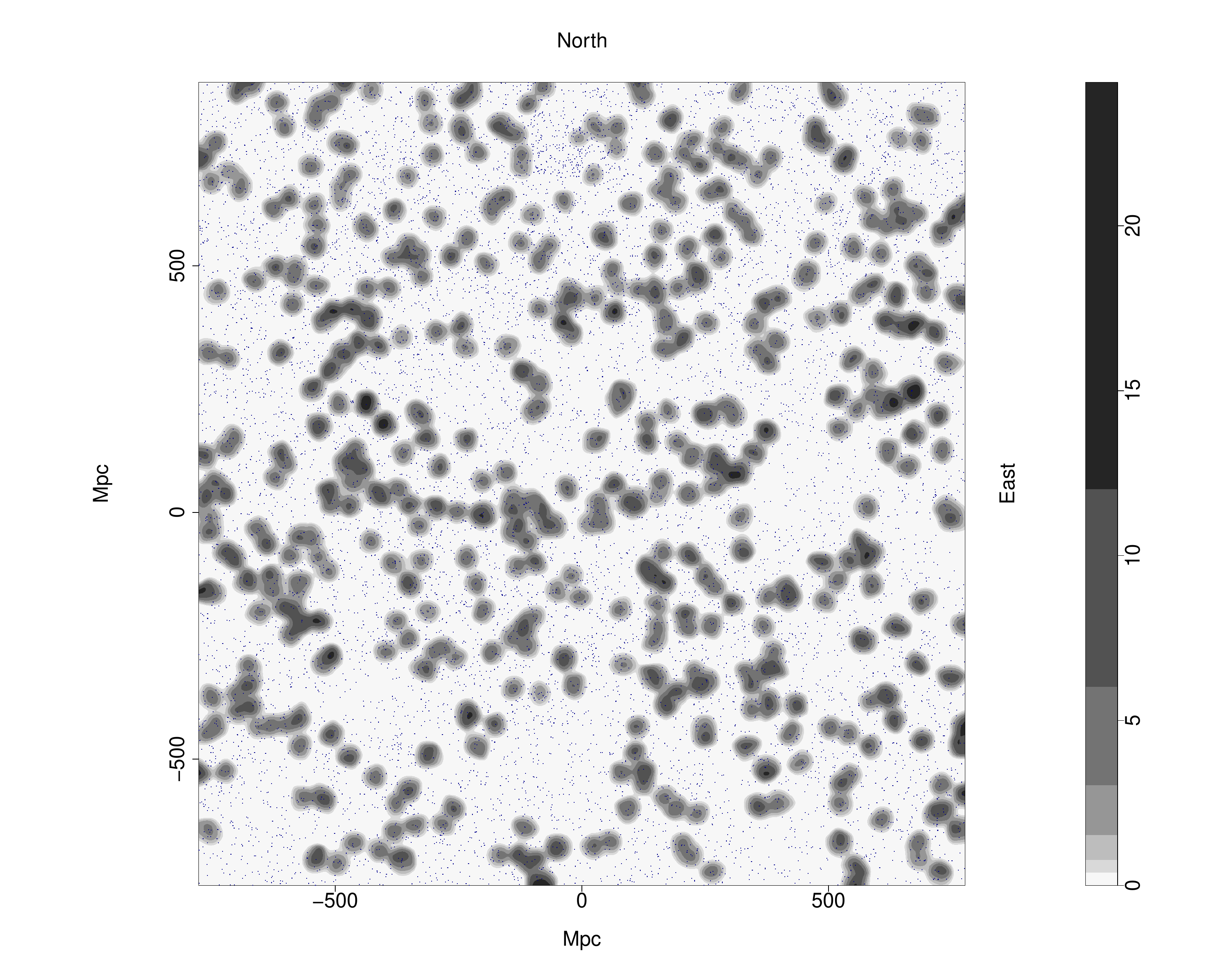}
        \caption{Density distribution of the flat-fielded Mg~{\sc II} absorbers 
          represented by the grey contours which have been smoothed using a 
          Gaussian kernel of $\sigma = 11$~Mpc and increase by a factor of two. 
          Blue dots represent the background probes (quasars). }
        \label{fig:CE_MgII_GA}
      \end{subfigure}
      \hfill
      \begin{subfigure}[b]{0.36\textwidth}
        \centering
        \includegraphics[width=\textwidth]{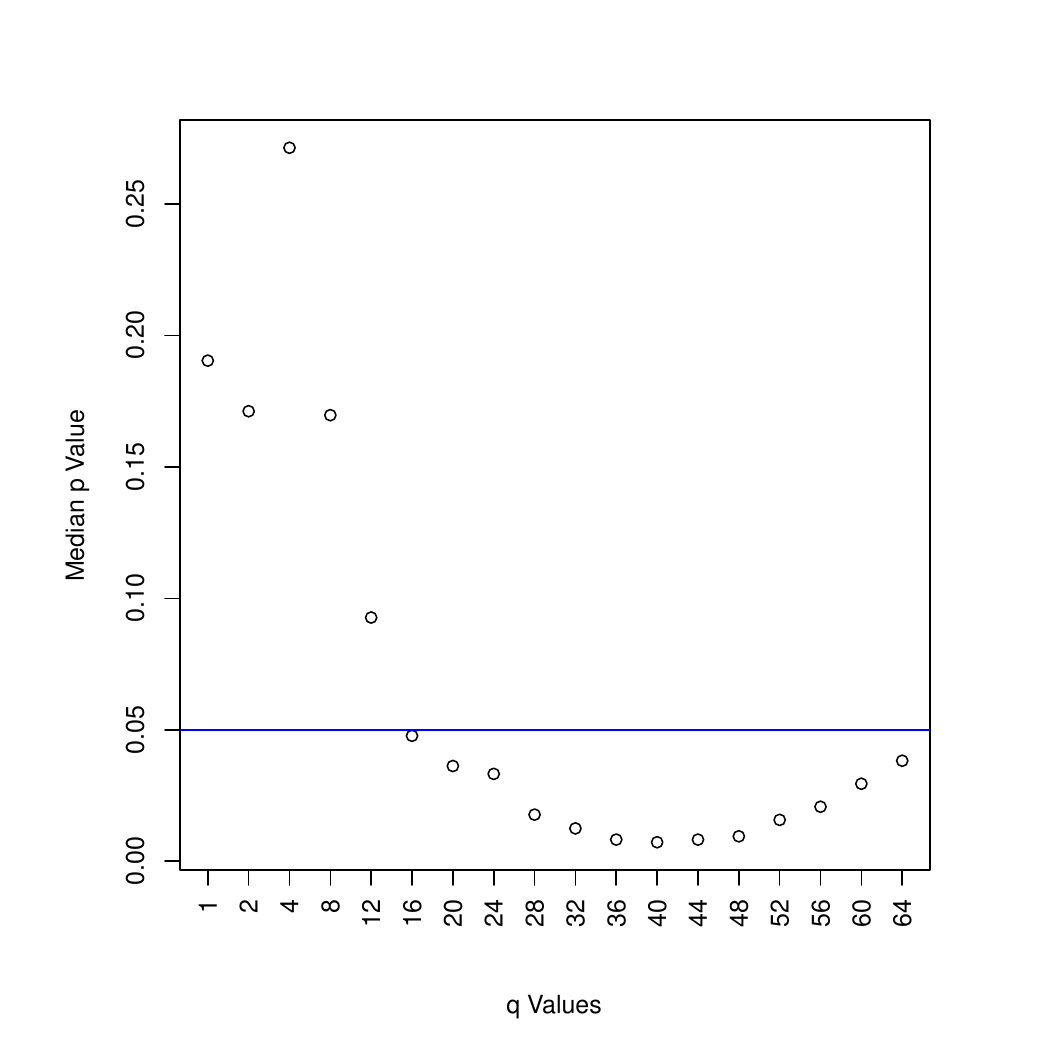}
        \caption{The median $p$-value over 100 runs of 2000 simulations as a
          function of chosen $q$ ($k$) value in the GA field; 
          see the adjacent image. The blue, 
          horizontal line is set to $p = 0.05$. The $p$-value is at a minimum
          of $0.0072$ when $q$ is $40$.}
        \label{fig:CE_med_GA}
      \end{subfigure}
        \caption{}
        \label{fig:CE_med_and_MgII_GA}
 \end{figure*}

\begin{figure*}
      \centering
      \begin{subfigure}[b]{0.5\textwidth}
        \centering
        \includegraphics[width=\textwidth]{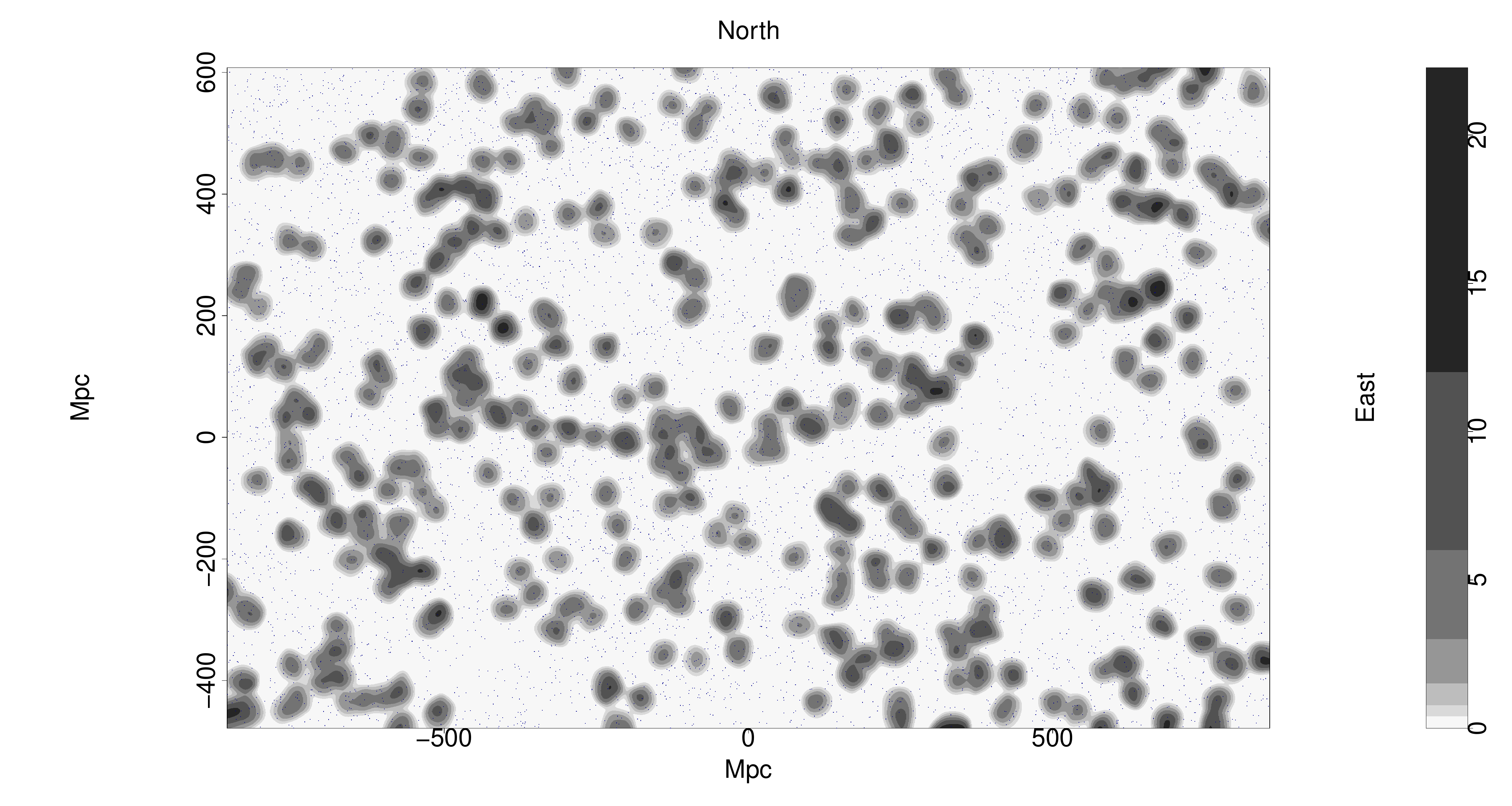}
        \caption{Density distribution of the flat-fielded Mg~{\sc II}
          absorbers represented by the grey contours which have been smoothed
          using a Gaussian kernel of $\sigma = 11$~Mpc and increase
          by a factor of two. Blue dots represent
          the background probes (quasars). This is the first `zoom' of the
          GA, where the GA field has been reduced in the north and south
          boundaries.}
        \label{fig:CE_MgII_GA_zoom1}
      \end{subfigure}
      \hfill
      \begin{subfigure}[b]{0.36\textwidth}
        \centering
        \includegraphics[width=\textwidth]{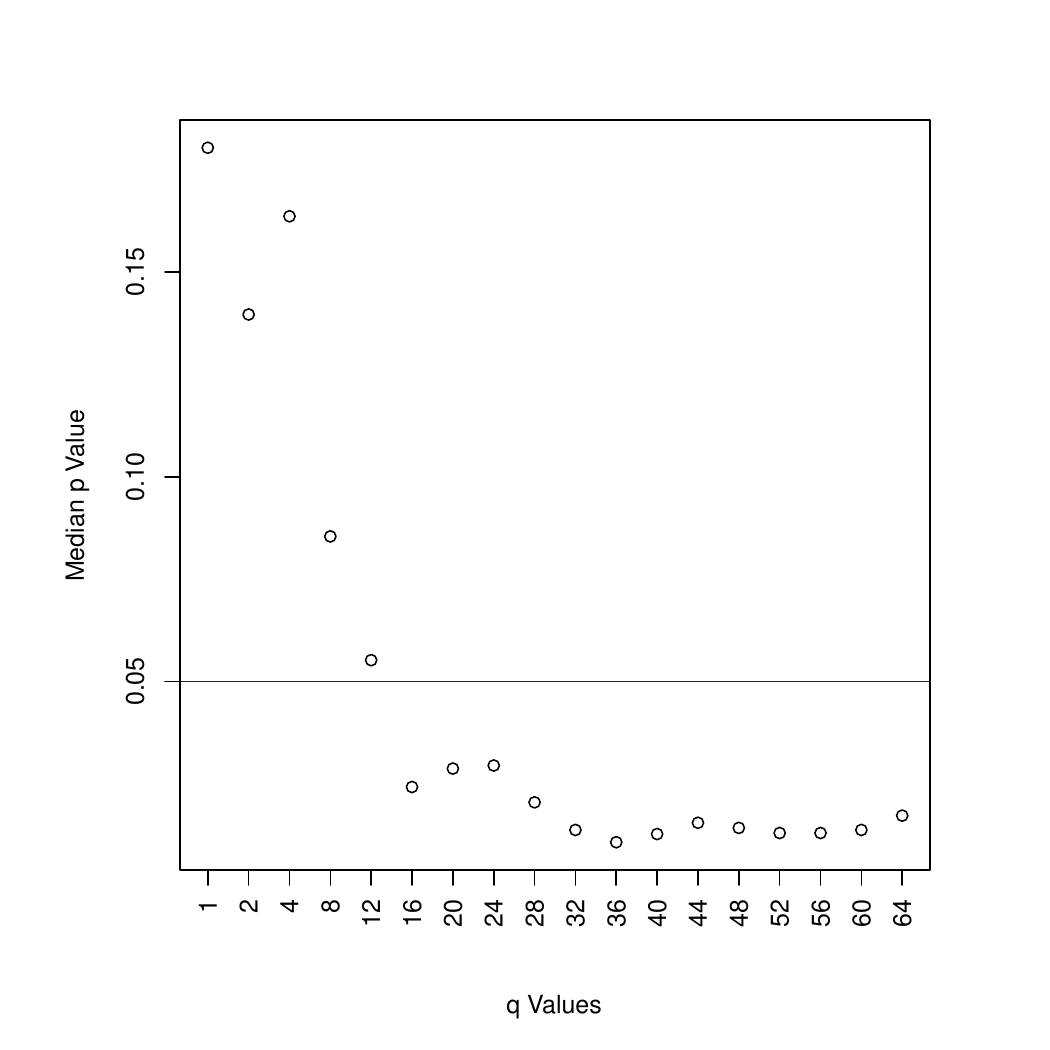}
        \caption{The median $p$-value over 100 runs of 2000 simulations as a
          function of chosen $q$ ($k$) value in the first `zoomed' field
          containing the GA; see the adjacent image. The blue, 
          horizontal line is set to $p = 0.05$. The $p$-value is at a minimum
          of $0.0107$ when $q$ is $36$.}
        \label{fig:CE_med_GA_zoom1}
      \end{subfigure}
        \caption{}
        \label{fig:CE_med_and_MgII_GA_zoom1}
 \end{figure*}

\begin{figure*}
      \centering
      \begin{subfigure}[b]{0.5\textwidth}
        \centering
        \includegraphics[width=\textwidth]{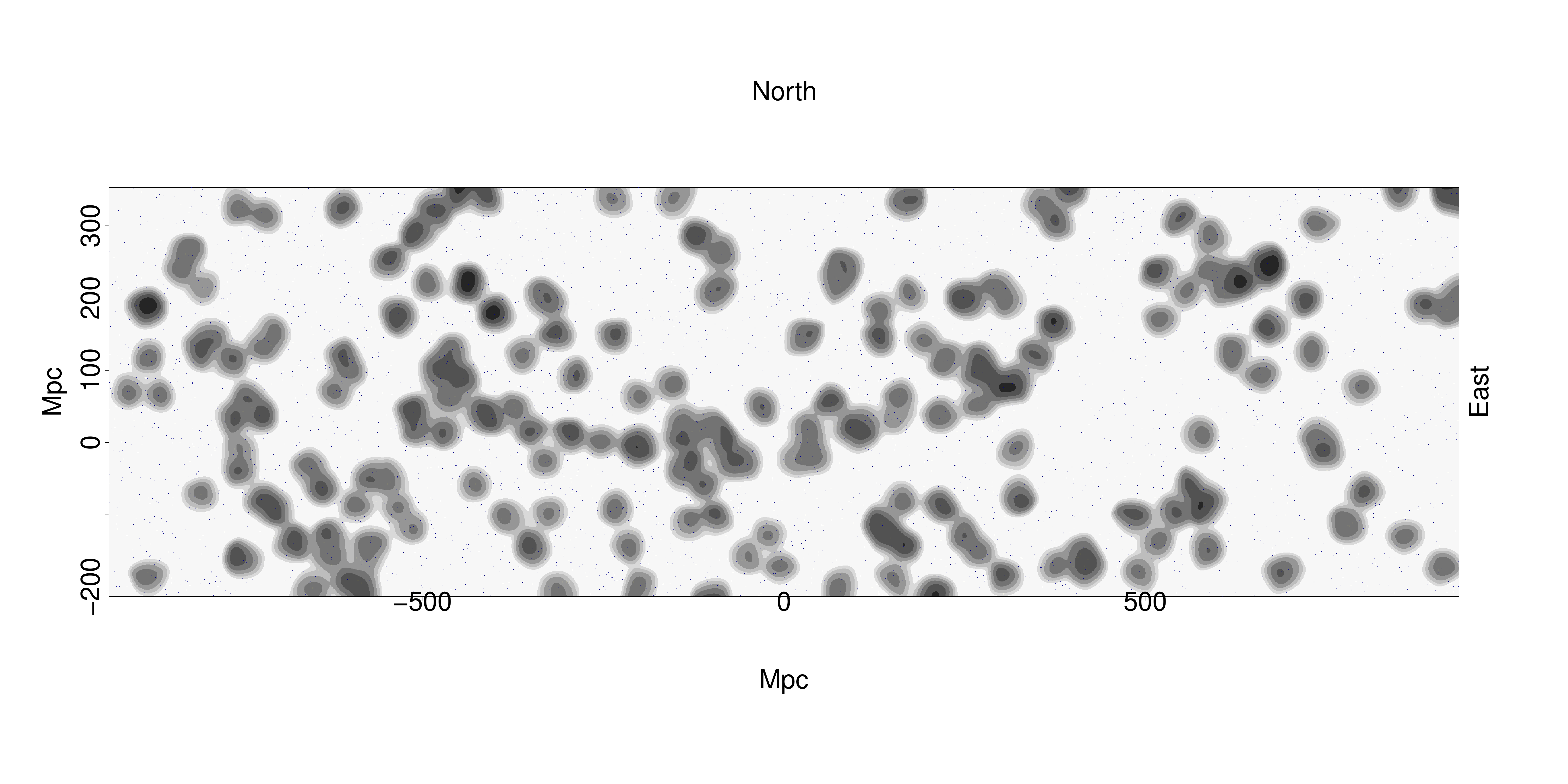}
        \caption{Density distribution of the flat-fielded Mg~{\sc II}
          absorbers represented by the grey contours which have been
          smoothed using a Gaussian kernel of $\sigma = 11$~Mpc and 
          increase by a factor of two. Blue dots
          represent the background probes (quasars). This is the second
          `zoom' of the GA, where the GA field has been further reduced in
          the north and south boundaries.}
        \label{fig:CE_MgII_GA_zoom2}
      \end{subfigure}
      \hfill
      \begin{subfigure}[b]{0.36\textwidth}
        \centering
        \includegraphics[width=\textwidth]{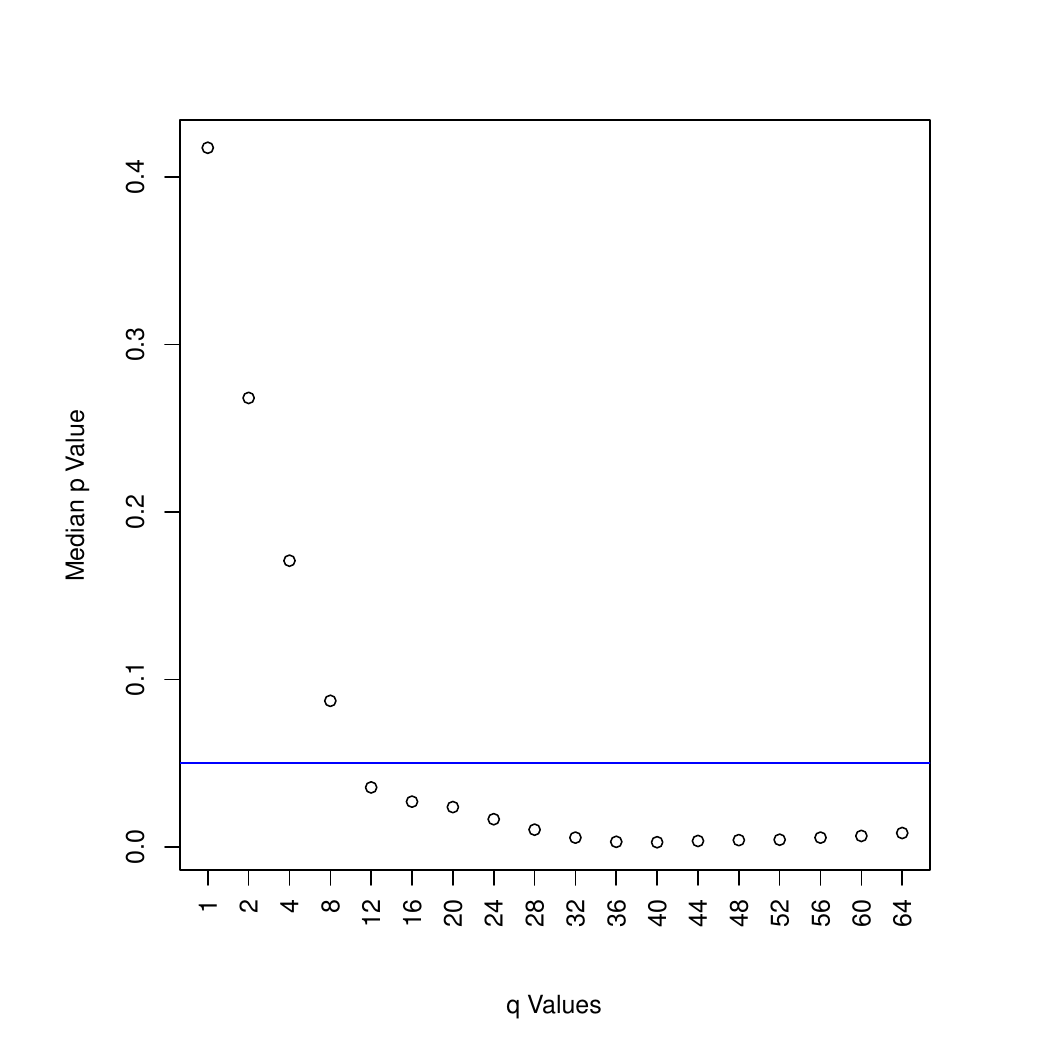}
        \caption{The median $p$-value over 100 runs of 2000 simulations as a
          function of chosen $q$ ($k$) value in the second `zoomed' field
          containing the GA; see the adjacent image. The blue, 
          horizontal line is set to $p = 0.05$. The $p$-value is at a minimum
          of $0.0027$ when $q$ is $40$.}
        \label{fig:CE_med_GA_zoom2}
      \end{subfigure}
        \caption{}
        \label{fig:CE_med_and_MgII_GA_zoom2}
 \end{figure*}

Then, to assess the (presumed) dominance of the GA itself we apply the test
to a succession of smaller fields (smaller in the north-south direction), all
centred on the GA. In Figs~\ref{fig:CE_med_and_MgII_GA} to 
\ref{fig:CE_med_and_MgII_GA_zoom2},
the median $p$-value over 100 runs of 2000 random simulations is shown
plotted against the chosen $q$ values, with the corresponding flat-fielded
Mg~{\sc II} image shown alongside.

The process of zooming into the GA allows the GA to become the dominating
feature in the field, which, as a result, increases the significance of
clustering (i.e.\ smaller $p$-value).  In the first Mg~{\sc II} field, Fig.
\ref{fig:CE_MgII_GA}, the minimum $p$-value is $0.0072$ at a $q$-value of
$40$, which (assuming a normal distribution) is equivalent to a significance
of $\sim 2.68\sigma$, Fig.~\ref{fig:CE_med_GA}. Whereas in the third Mg~{\sc
  II} field, Fig.~\ref{fig:CE_MgII_GA_zoom2}, the minimum $p$-value drops to
$0.0027$ at a $q$-value of 40, which is equivalent to a significance of
$3.00\sigma$, Fig.~\ref{fig:CE_med_GA_zoom2}. In this way we can judge that
the GA is the dominant, contributing factor to the significant level of
clustering in the field.

\begin{figure*}
      \centering
      \begin{subfigure}[b]{0.5\textwidth}
        \centering
        \includegraphics[width=\textwidth]{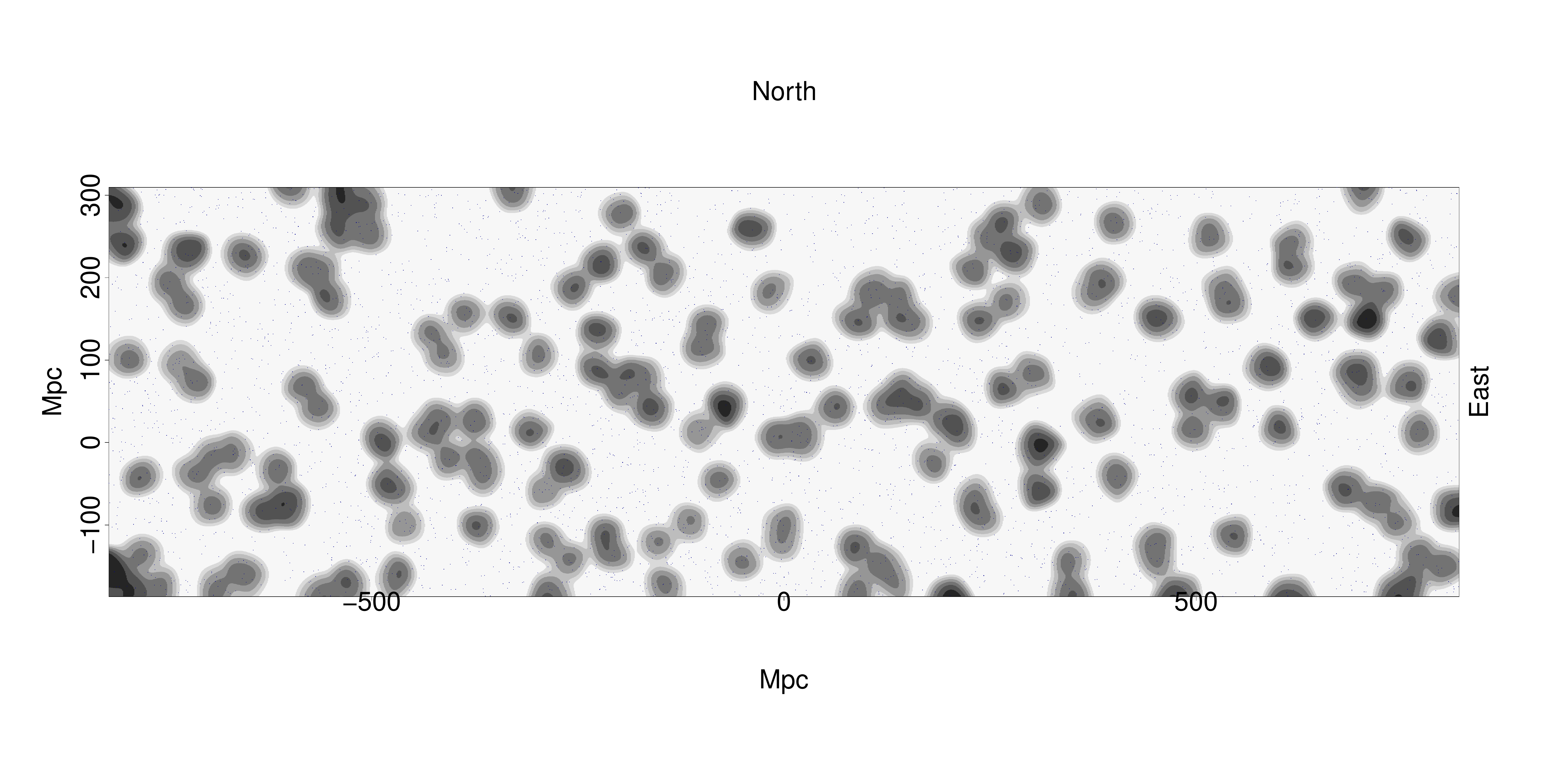}
        \caption{Density distribution of the flat-fielded Mg~{\sc II}
          absorbers in the redshift interval $z = 0.682 \pm 0.060$, on the
          same sky coordinates of the `zoomed' GA field, represented by the
          grey contours which have been smoothed using a Gaussian kernel of
          $\sigma = 11$~Mpc and increase by a factor of two. Blue dots
          represent the background probes (quasars).}
        \label{fig:CE_MgII_zoom2_0_682}
      \end{subfigure}
      \hfill
      \begin{subfigure}[b]{0.36\textwidth}
        \centering
        \includegraphics[width=\textwidth]{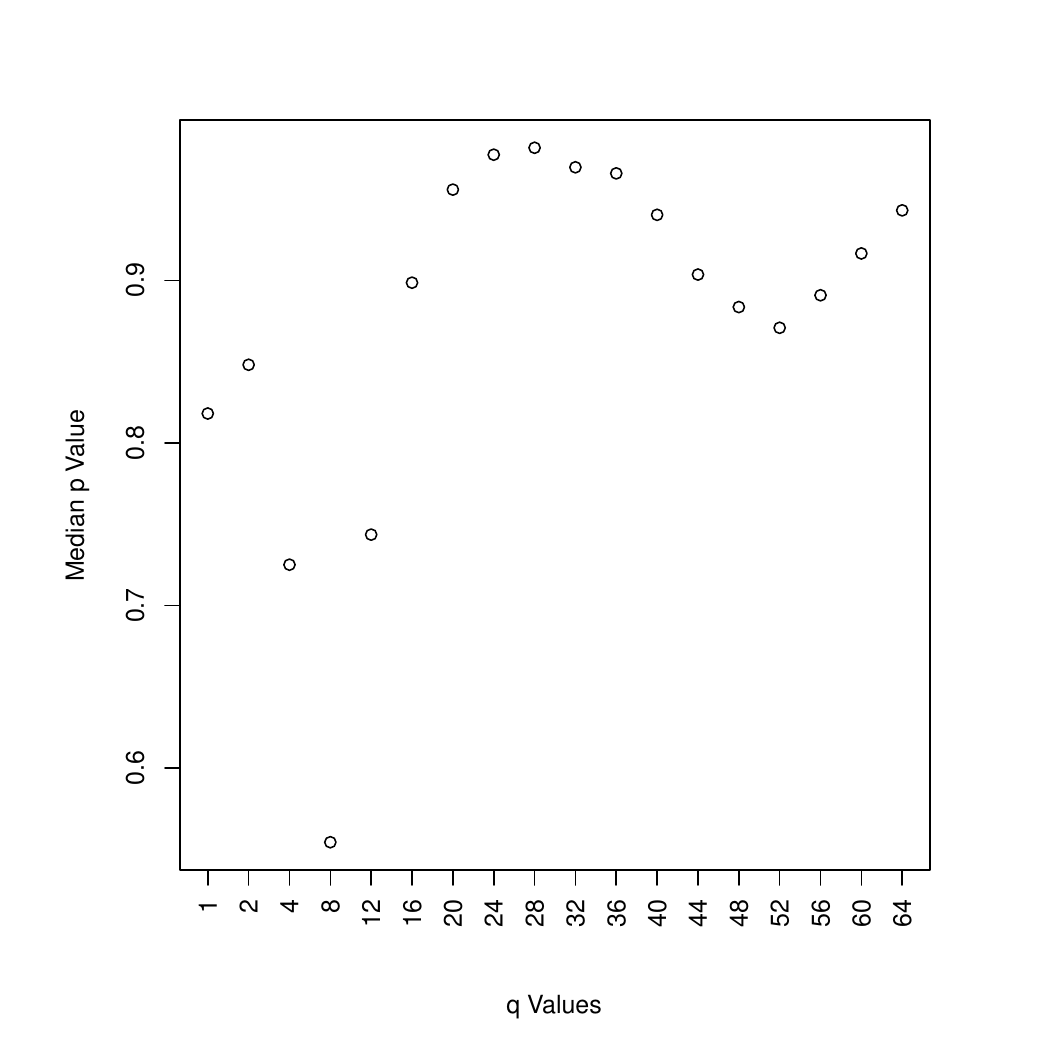}
        \caption{The median $p$-value over 100 runs of 2000 simulations as a
          function of chosen $q$ ($k$) value in the $z=0.682 \pm 0.060$
          redshift interval on the same sky coordinates as the `zoomed' GA
          field; see the adjacent image.}
        \label{fig:CE_med_zoom2_0_682}
      \end{subfigure}
        \caption{}
        \label{fig:CE_med_and_MgII_zoom2_0_682}
 \end{figure*}

\begin{figure*}
      \centering
      \begin{subfigure}[b]{0.5\textwidth}
        \centering
        \includegraphics[width=\textwidth]{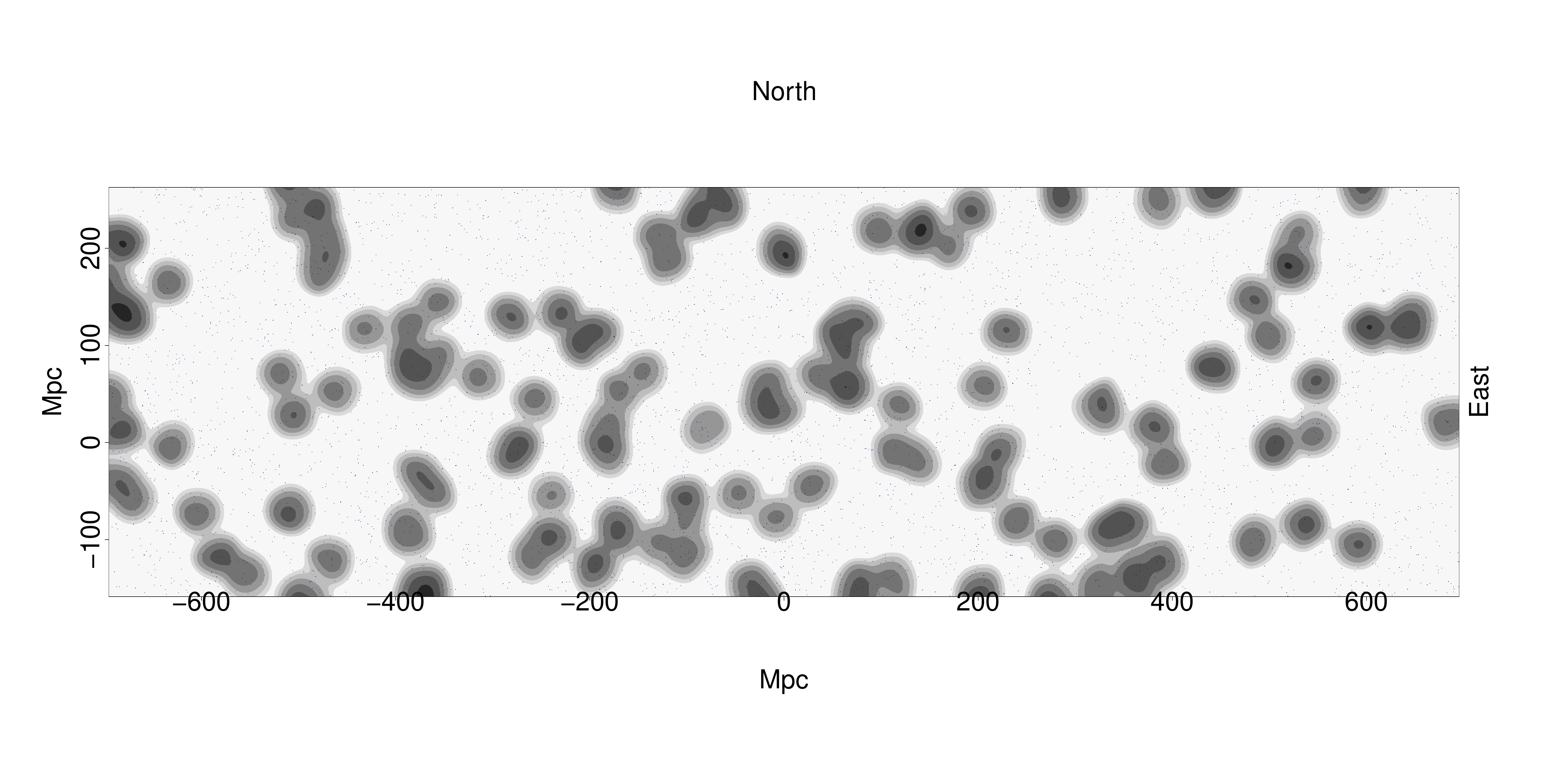}
        \caption{Density distribution of the flat-fielded Mg~{\sc II}
          absorbers in the redshift interval $z = 0.562 \pm 0.060$, on the
          same sky coordinates of the `zoomed' GA field, represented by the
          grey contours which have been smoothed using a Gaussian kernel of
          $\sigma = 11$~Mpc and increase by a factor of two.  Blue dots
          represent the background probes (quasars). }
        \label{fig:CE_MgII_zoom2_0_562}
      \end{subfigure}
      \hfill
      \begin{subfigure}[b]{0.36\textwidth}
        \centering
        \includegraphics[width=\textwidth]{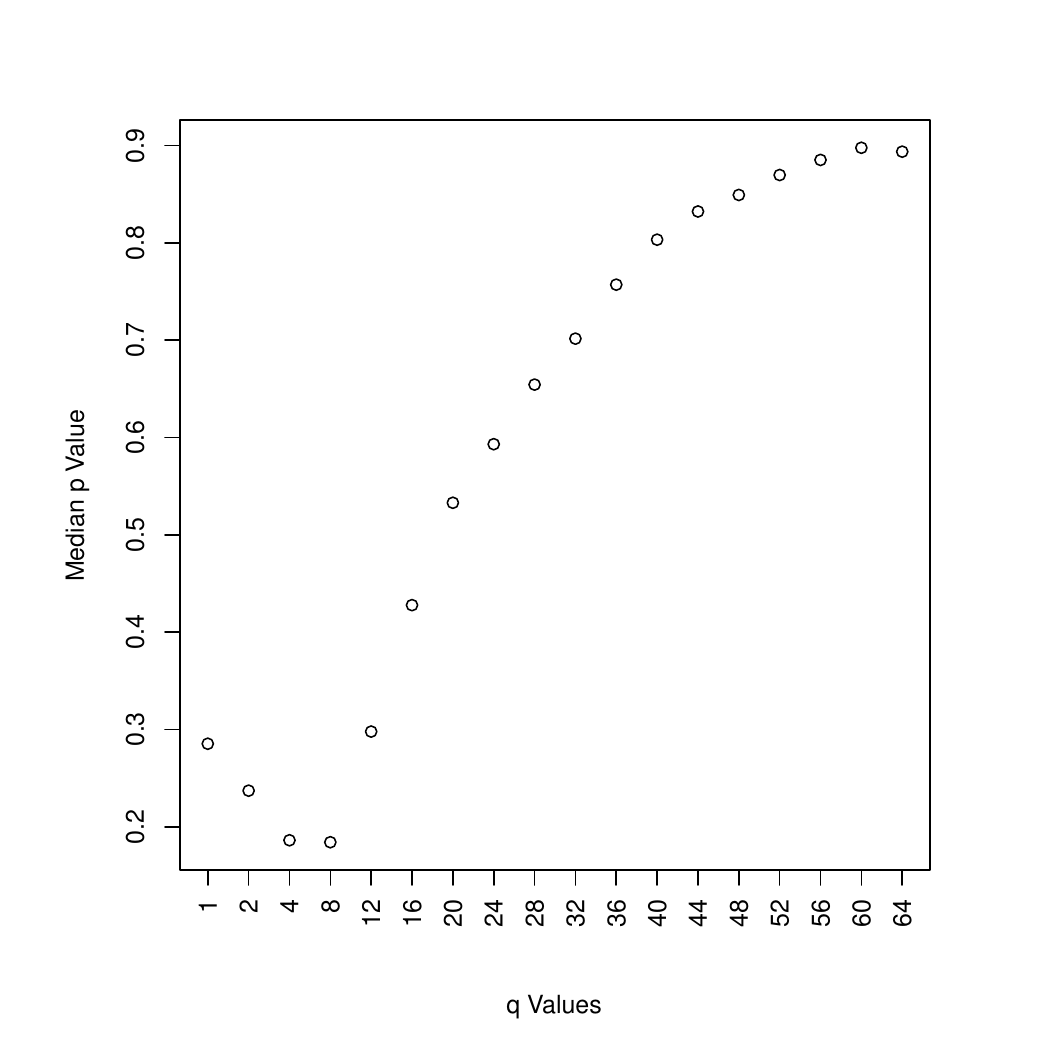}
        \caption{The median $p$-value over 100 runs of 2000 simulations as a
          function of chosen $q$ ($k$) value in the $z=0.562 \pm 0.060$
          redshift interval on the same sky coordinates as the `zoomed' GA
          field; see the adjacent image.}
        \label{fig:CE_med_zoom2_0_562}
      \end{subfigure}
        \caption{}
        \label{fig:CE_med_and_MgII_zoom2_0_562}
 \end{figure*}

\begin{figure*}
      \centering
      \begin{subfigure}[b]{0.5\textwidth}
        \centering
        \includegraphics[width=\textwidth]{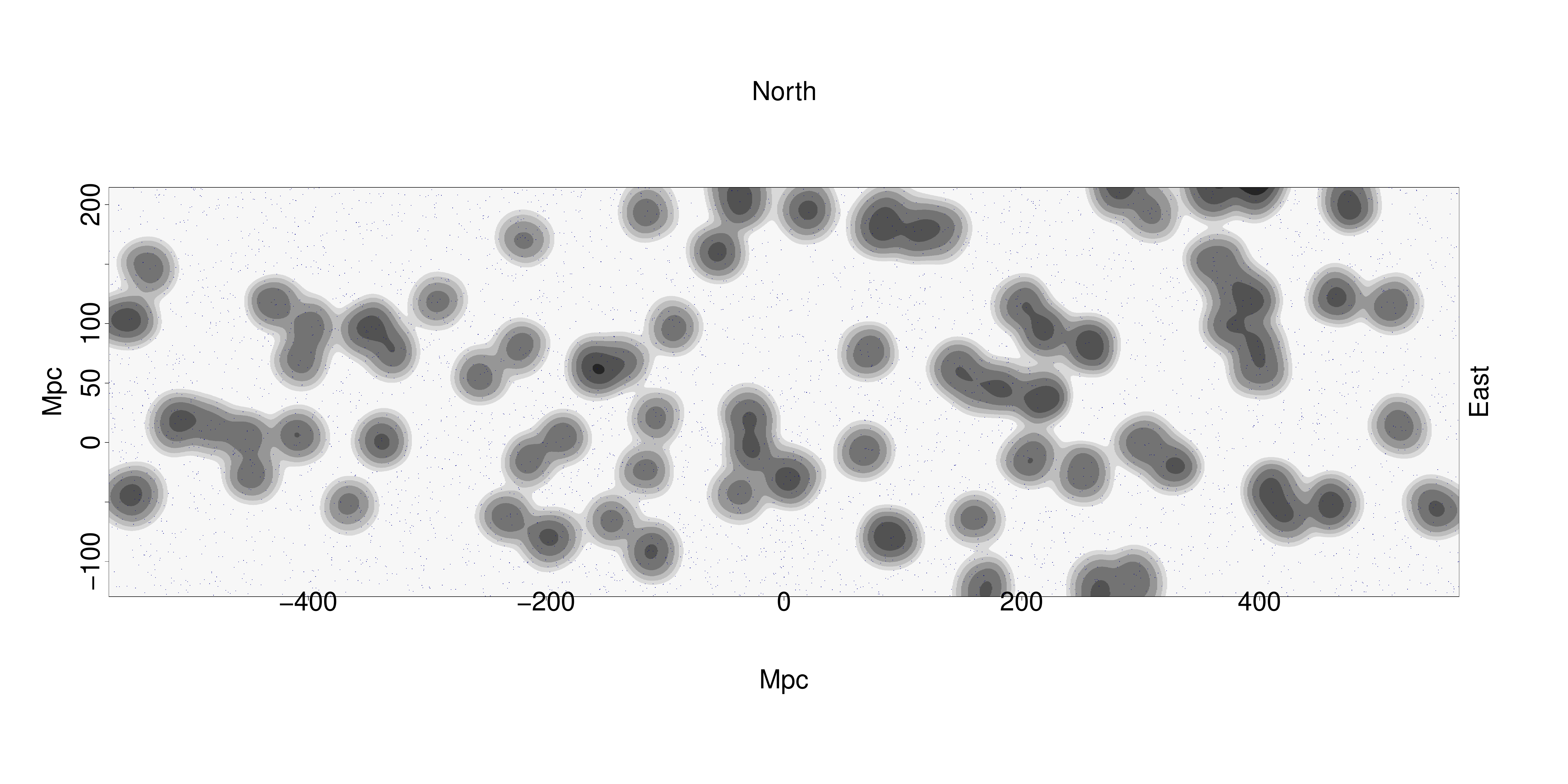}
        \caption{Density distribution of the flat-fielded Mg~{\sc II}
          absorbers in the redshift interval $z = 0.446 \pm 0.060$, on the
          same sky coordinates of the `zoomed' GA field, represented by the
          grey contours which have been smoothed using a Gaussian kernel of
          $\sigma = 11$~Mpc and increase by a factor of two. Blue dots
          represent the background probes (quasars).}
        \label{fig:CE_MgII_zoom2_0_446}
      \end{subfigure}
      \hfill
      \begin{subfigure}[b]{0.36\textwidth}
        \centering
        \includegraphics[width=\textwidth]{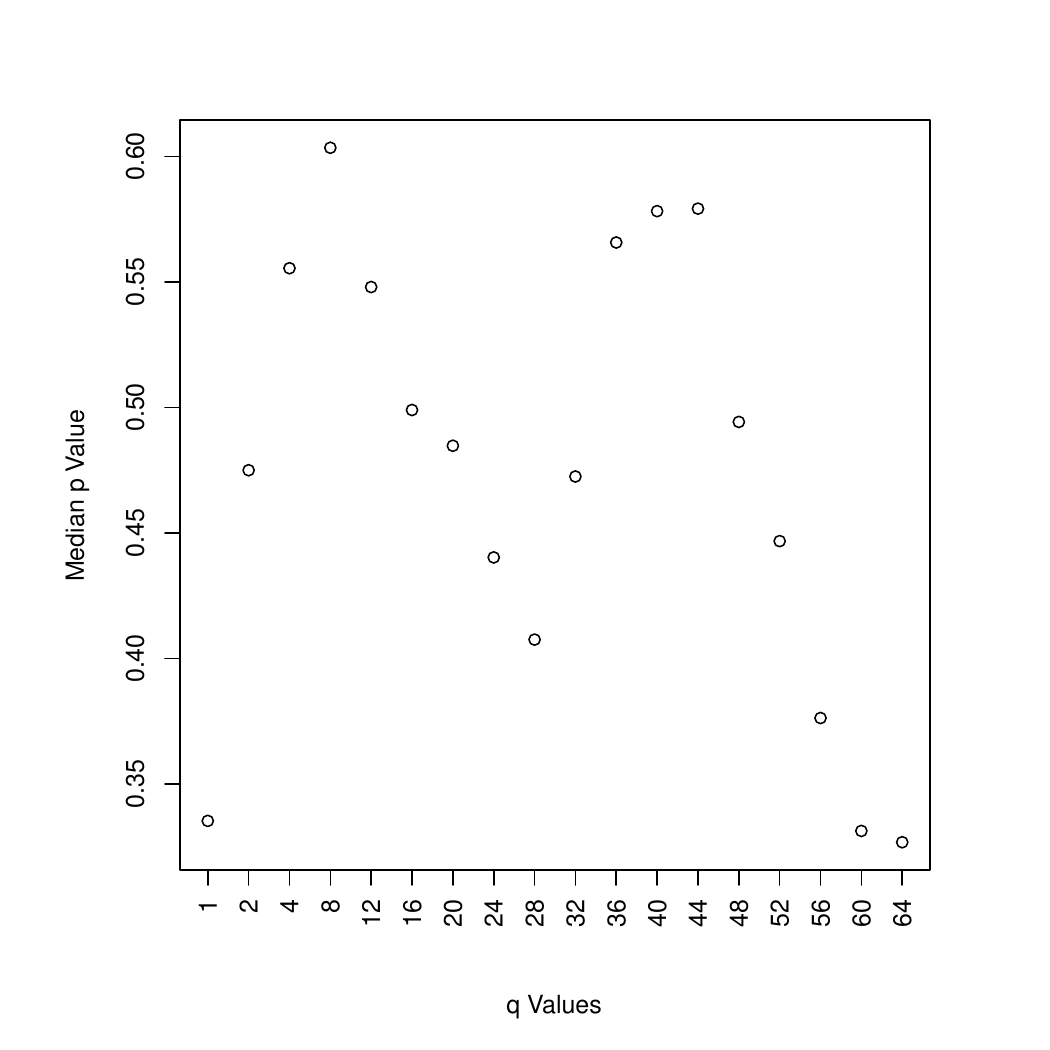}
        \caption{The median $p$-value over 100 runs of 2000 simulations as a
          function of chosen $q$ ($k$) value in the $z=0.446 \pm 0.060$
          redshift interval on the same sky coordinates as the `zoomed' GA
          field; see the adjacent image.}
        \label{fig:CE_med_zoom2_0_446}
      \end{subfigure}
        \caption{}
        \label{fig:CE_med_and_MgII_zoom2_0_446}
 \end{figure*}

The heuristic process of `zooming' into the GA was next applied to three
other fields at lower redshift slices ($z$: $0.682$, $0.562$, $0.442$)
centred on the sky coordinates of the GA. The background probes are kept the
same in the three new fields as those in the GA field, allowing a direct
comparison of clustering in just the Mg~{\sc II} absorbers (as
  in Section \ref{subsect:MST}).  Figs~\ref{fig:CE_med_and_MgII_zoom2_0_682}
to \ref{fig:CE_med_and_MgII_zoom2_0_446} show the results of the CE test for
the three lower redshift fields using the smallest field size (i.e.\ the
second `zoom'). The $p$-value profiles as a function of $q$-value in each of
the lower redshift fields appear more scattered and varied compared with the
GA results.

The major difference between the $p$-value profiles for the GA field and the
$p$-value profiles for three lower redshift fields is that there is no sign
of any significant results ($p$-value $< 0.05$) in any of the three lower
redshift fields for any of the chosen $q$-values.  The background probes were
the same in all four fields (GA field and the three lower redshift fields)
indicating that the Mg~{\sc II} absorbers are responsible for the different
$p$-value profiles. From this we can assert that the GA field is markedly
distinct, with significant clustering attributable to the GA.

As a further test of the dominance of the GA in the CE statistics we have
applied our `polygon-approach'. Visually, we draw a polygon around what we
identify visually as the member absorbers of the GA. We leave the absorbers
in the polygon untouched but reassign (i.e.\ shuffle) at random the
$y$-coordinates of absorbers outside the polygon, while avoiding the area
within the polygon. We apply this process to the data of
Fig.~\ref{fig:CE_MgII_GA_zoom2}. In this way we can compare the CE statistics
arising from the original data with those in which the GA points inside the
polygon are unchanged but those outside the GA polygon are randomised. We
find that the range of $p$-values for the original data ($p$-values $\sim$
0.002--0.003) is very similar to that of the GA $+$ randomised data
($p$-values $\sim$ 0.001--0.003), suggesting that the GA is indeed the
dominant source of the clustering signal.

\subsubsection{Power Spectrum Analysis}

Power Spectrum Analysis (PSA) --- see mainly \citet{Webster1976a}, but also
\citet{Webster1976b} and \citet{Webster1982} --- is a powerful Fourier method
for assessing the presence and significance of clustering in rectangular (2D
PSA) or cuboidal fields (3D PSA). PSA was designed to be effective for the
detection of clustering that may be weak and escape detection by other
methods; it is, however, not a case-control method. A brief summary of the
theory of PSA may be found in section~5 of \citet{Clowes1986}.

\begin{figure}
    \centering
    \includegraphics[scale=0.65]{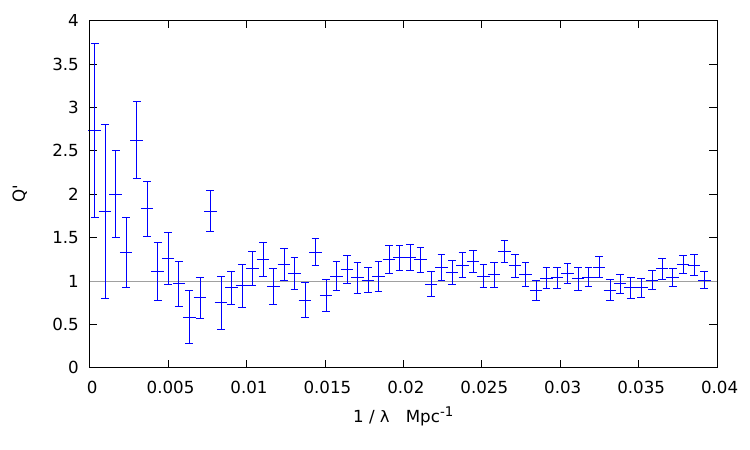}
    \caption{The plot shows the PSA statistic $Q'$ plotted against $1 /
      \lambda$, with $\lambda$ in Mpc, for the 2D PSA. The bin size is $6.7
      \times 10^{-4}\,\mathrm{Mpc^{-1}}$ and the error bars are $\pm
      \sigma$. The horizontal line $Q' = 1$ indicates the expectation value
      in the case of no clustering. The (six) high points towards the left of
      the plot allow a clustering scale of $\lambda_c \sim 270\,\mathrm{Mpc}$
      to be identified. The final PSA statistic $Q$ for this scale
      $\lambda_c$ corresponds to a detection of clustering at a significance
      of $4.8\sigma$.}
    \label{fig:2dPSA_1}
\end{figure}

We apply 2D PSA to the same rectangular field, illustrated in
Fig.~\ref{fig:CE_MgII_GA_zoom2}, that was used for the CE analysis above.
Fig.~\ref{fig:2dPSA_1} shows the plot of the intermediate PSA statistic $Q'$
against $1 / \lambda$. The (six) high points towards the left of the plot
allow a clustering scale of $\lambda_c \sim 270\,\mathrm{Mpc}$ to be
identified. The final PSA statistic $Q$ for this scale $\lambda_c$
corresponds to a detection of clustering at a significance of $4.8\sigma$.

We have applied the polygon-approach here also. As in the discussion above
for the CE method, we leave the GA absorbers in the polygon untouched but
reassign (i.e.\ shuffle) at random the $y$-coordinates of absorbers outside
the polygon, while avoiding the area within the polygon. In this way we can
establish that the significance from the 2D PSA of the GA absorbers {\it
  alone} --- i.e.\ with other absorbers randomised --- has a mean value $\sim
3.5\sigma$ (with a range 3.0--4.4$\sigma$) for the original scale $\lambda_c
\sim 270\,\mathrm{Mpc}$. In fact, the value of $\lambda_c$ for the
polygon-approach varies too, with the mean significance at the actual values
of $\lambda_c$ being $\sim 3.8\sigma$.

From this polygon-approach, it appears likely that, while the GA is the
dominant contributor to the PSA result, a smaller contribution from other
absorbers in the field is detected too. This outcome might be expected, given
the power of the PSA method. The failure to detect a contribution from the
other absorbers with the CE method could be because the CE method is
intrinsically less sensitive, or because its case-control correction has
successfully eliminated artefacts from the background probes (the controls).

The polygon-approach can also be used to assess the relative power of the 2D
PSA and the CE test. For example, we reduced the number of GA absorbers in
the polygon from 52 to 42 by random selection (with the points outside the
polygon being randomised as usual but unchanged in total number). In that
case, the GA is generally not detected by the CE test, at a significance
level of 0.01 ($2.3\sigma$), but is generally still detected by the PSA, at
$> 2.7\sigma$. The power of a statistical test to discriminate is an
important factor, and an uninteresting $p$-value does not necessarily mean
nothing interesting in the data. \citet{Webster1976a} demonstrates that the
PSA has more power to detect clustering than a simple nearest-neighbour test.
The CE test uses multiple neighbours, and so can be expected to have more
power than a nearest-neighbour test, but, as our tests with the
polygon-approach suggest, still has less power than the PSA. Of course, the
CE test has the useful feature of case-control comparisons, whereas the PSA
does not.

\subsection{Overdensity}

As seen previously, the SLHC / CHMS method splits the GA into two
agglomerations - one large, statistically significant portion which makes up
the majority of what we visually identified as the GA, and one smaller,
statistically not significant portion which makes up the remainder of what we
visually identified as part of the GA. We will refer to these agglomerations
as GA-main and GA-sub, respectively, for simplicity.  The overdensity of the
GA can be calculated using the CHMS approach as described earlier in the
paper (section \ref{subsect:MST}).  However, in the case of a strongly curved
structure such as the GA (and GA-main), the MST-based method of
\citet{Pilipenko2007} can have some advantages. We shall refer to these two
methods as CHMS-overdensity and MST-overdensity, respectively. The
MST-overdensity does not consider the physical volume of the structure being
assessed. Instead, it calculates the overdensity based on the MST
edge-lengths: $\delta = \langle l_0^3 \rangle / \langle l^3 \rangle - 1$
where $l$ is MST edge-length for the structure and $l_0$ is that for a
control field.  Given the curvature of GA-main, the CHMS volume and
CHMS-overdensity refer to a volume that encloses both GA-main and some space
above it (where there are rather fewer absorbers, and those are not related
to the GA). Therefore, the CHMS method is likely to overestimate the volume
and underestimate the overdensity. In contrast, the MST-overdensity, which is
an internal measure that considers only the points belonging to the group,
and no additional space arising from curvature, is likely to be a better
estimate of the overdensity.  Conversely, GA-sub is a globular shape, so it
is possible to construct a unique volume enclosing only the absorbers
attached to GA-sub and not additional ones at lower density.  Therefore, the
CHMS-overdensity calculation for GA-sub is likely to be a fair estimate.

GA-main, as mentioned earlier, has a significance of $4.5\sigma$, while
GA-sub has a much smaller significance of $2.3\sigma$.  By splitting the
usual control field (using the {\it larger} field-of-view, see Fig.
\ref{fig:qso_bigger_FOV}) into eight portions --- four quarter segments and
four half segments --- we repeatedly calculate the significance and the
overdensities of GA-main and GA-sub using different control fields.  An
uncertainty can then be estimated for the significance and overdensities for
both portions of the GA.  Our results are as follows: (1) GA-main, containing
44 Mg~{\sc II} absorbers, has a significance of $(4.5 \pm 0.6)\sigma$; a
CHMS-overdensity of $\delta \rho_{CHMS} / \rho_{CHMS} = 0.9 \pm 0.6$; and an
MST-overdensity of $\delta \rho_{MST} / \rho_{MST} = 1.3 \pm 0.3$; (2)
GA-sub, containing 11 Mg~{\sc II} absorbers, has a significance of $(2.1 \pm
0.9)\sigma$; a CHMS-overdensity of $\delta \rho_{CHMS} / \rho_{CHMS} = 1.5
\pm 0.3$; and an MST-overdensity of $\delta \rho / \rho = 1.3 \pm 0.3$.  As
expected, the CHMS-overdensity is lower than the MST-overdensity for GA-main,
indicating that the CHMS unique volume encapsulating GA-main is likely to be
an overestimate because of the curvature of the arc. In contrast, for
GA-sub, which has a globular-shape, the CHMS-overdensity and the
MST-overdensity have similar values, as expected when there is no marked
curvature.  In addition, the CHMS-overdensity has a much larger error than
the MST-overdensity which suggests giving preference to the latter.  Notice
here that both GA-main and GA-sub have the same MST-overdensity, which
supports their belonging to the same structure.

A final method of calculating the number overdensity is to simply draw a
rectangle around the visually-selected Mg~{\sc II} absorbers in the GA and
compare the number of absorbers per unit area in the rectangle to the number
of absorbers in the whole field. The method will underestimate the GA
overdensity for three reasons: (i) the GA contributes to the density of the
whole field, although only by a small fraction; (ii) the rectangular shape
around the GA overestimates the area encompassing the GA as the GA is curved,
therefore having a large portion of `empty' space (with non-GA absorber
members); (iii) this method encompasses the {\it whole} GA from visual
inspection, rather than separating it into two agglomerations like the CHMS
method, thus reducing the overall density in the GA rectangle.  Using this
method we calculate an overdensity of $\delta \rho / \rho \sim 0.93$.

In addition to the number overdensity, we can estimate the mass excess by
assuming $\delta_n = \delta_m$, where $\delta_n$ is the MST-overdensity and
$\delta_m$ is the mass overdensity.  We use here the MST-overdensity, rather
than the CHMS-overdensity.  We are here taking the critical density of the
universe to be $9.2 \times 10^{-27}\,\mathrm{kg\,m^{-3}}$, as calculated
using the cosmological parameters used throughout this paper, and the
matter-energy density parameter to be $\Omega_{M0} = 0.27$. The mass excesses
for GA-main and GA-sub are then $1.8 \times 10^{18} M_\odot$ and $3.4 \times
10^{17} M_\odot$ respectively. Note that the mass excess of GA-main $+$
GA-sub is comparable to that of the Huge-LQG \citep{Clowes2013}.

\subsection{Comparisons with other data}

Independent corroboration of a very large LSS by an independent tracer can
provide compelling support. In the case of the Huge-LQG \citep{Clowes2013}, a
$\sim$ Gpc structure of quasars, independent corroboration was provided by
Mg~{\sc II} absorbers. Here, we can invert this approach and look for
corroboration of the GA, a $\sim$ Gpc structure of Mg~{\sc II} absorbers, in
quasars. We use the SDSS DR16Q database \citep{Lyke2020}. In addition, we
look at the databases of DESI galaxy clusters from \citet{Zou2021}.

We are concerned at this stage with simple visual inspection, and will leave
the subtleties of correcting for possible artefacts in the DR16Q quasars and
the DESI clusters to future work. Our approach here will be simply to
superimpose contours for the spatial distribution of the quasars (in blue)
and the clusters (in green) onto the Mg~{\sc II} density images (grey, as
previously).

\begin{figure*}
      \centering
      \begin{subfigure}[b]{0.45\textwidth}
        \centering
        \includegraphics[width=\textwidth]{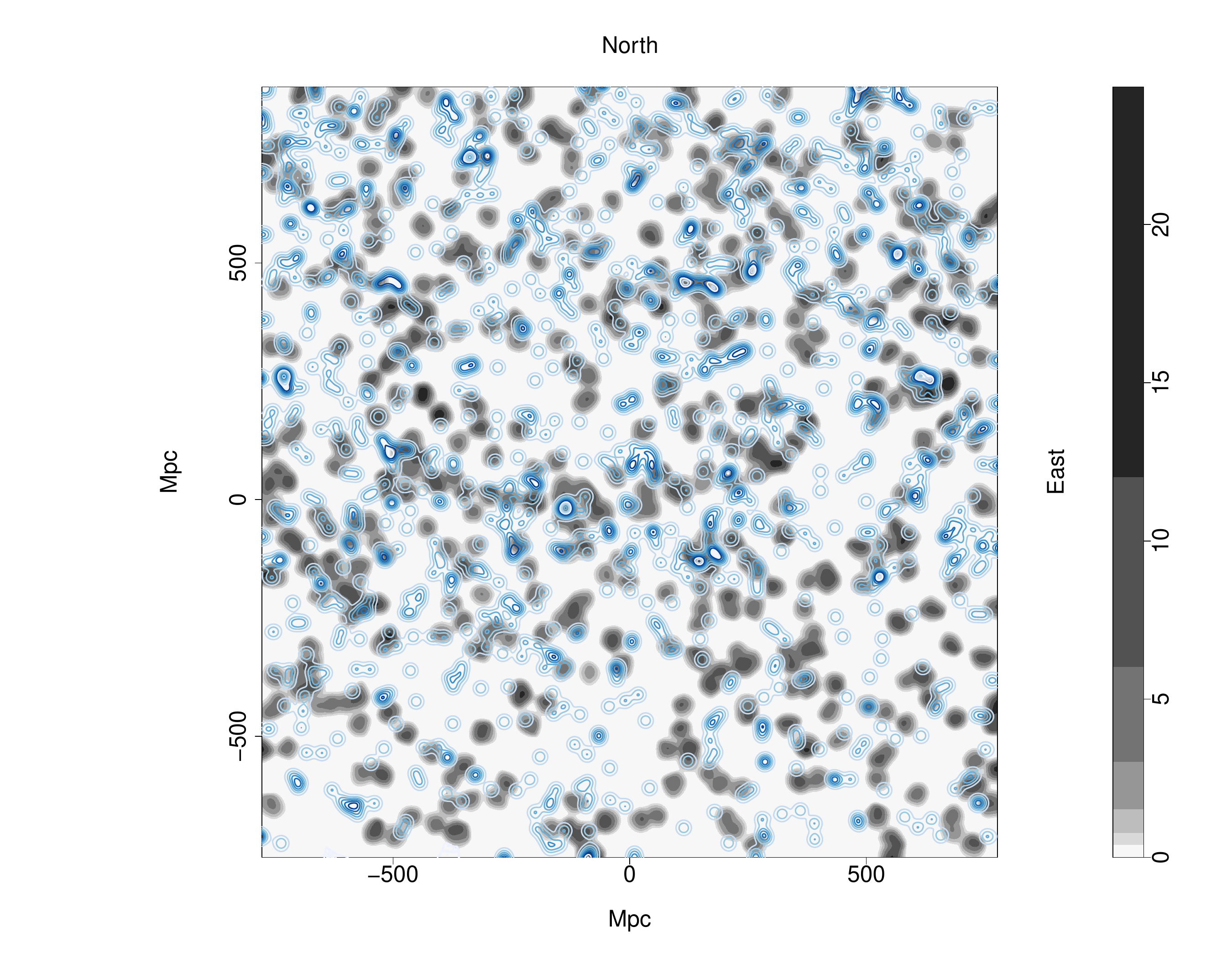}
        \caption{Density distribution of the flat-fielded Mg~{\sc II}
          absorbers in the redshift slice $z=0.802 \pm 0.060$ represented by
          the grey contours which have been smoothed using a Gaussian kernel
          of $\sigma = 11$~Mpc and increase by a factor of two. Blue
          contours represent the DR16Q quasars with $i \le 20.0$, in the same
          redshift slice as the Mg~{\sc II} absorbers, smoothed using a
          Gaussian kernel of $\sigma = 11$~Mpc and increasing by a factor of
          two.  The GA can be seen stretching across $\sim 1$ Gpc in the
          centre of the figure (at tangent-plane $y$-coordinate $\sim
          0$~Mpc). Visually, the blue contours can be seen to follow the same
          general trajectory as the grey contours, indicating an association
          between the Mg~{\sc II} absorbers and the DR16Q quasars.}
        \label{fig:MgII_and_dr16_quasars_imag_20}
       \end{subfigure}
       \hfill
       \begin{subfigure}[b]{0.45\textwidth}
        \centering
        \includegraphics[width=\textwidth]{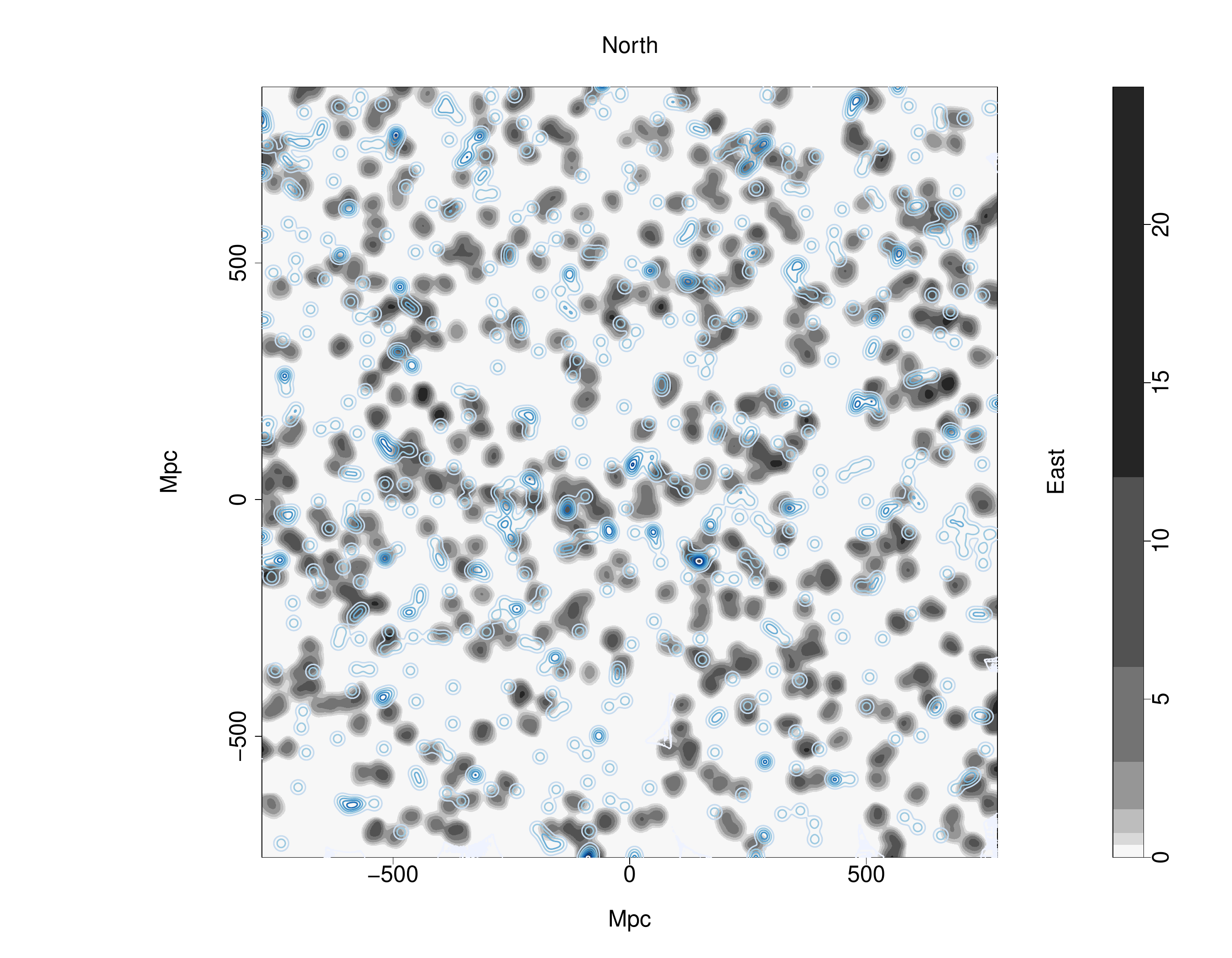}
        \caption{Density distribution of the flat-fielded Mg~{\sc II}
          absorbers in the redshift slice $z = 0.802 \pm 0.060$ represented
          by the grey contours which have been smoothed using a Gaussian
          kernel of $\sigma = 11$~Mpc and increase by a factor of two. Blue
          contours represent the DR16Q quasars, with $i \leq 19.5$, in the
          same redshift slice as the Mg~{\sc II} absorbers, smoothed using a
          Gaussian kernel of $\sigma = 11$~Mpc and increasing by a factor of
          two.  The GA can be seen stretching across $\sim 1$~Gpc in the
          centre of the figure (at tangent-plane $y$-coordinate $\sim
          0$~Mpc). Visually, the blue contours can be seen to follow the same
          general trajectory as the grey contours, indicating an association
          between the Mg~{\sc II} absorbers and the DR16Q quasars.}
        \label{fig:MgII_and_dr16_quasars_imag_19_5}
      \end{subfigure}
        \caption{}
        \label{fig:MgII_and_dr16_quasars}
 \end{figure*}

We begin with the quasars, selected for the same redshift interval as the GA
--- Fig.~\ref{fig:MgII_and_dr16_quasars}. We show two cases, one for quasars
with $i \le 20.0$ (Fig.~\ref{fig:MgII_and_dr16_quasars_imag_20}) and one for
$i \le 19.5$ (Fig.~\ref{fig:MgII_and_dr16_quasars_imag_19_5}). We anticipate
that we should then be restricting to `traditional' high-luminosity
quasars. In both cases, it is immediately clear that the quasars follow the
same general trajectory as the GA. The quasars are entirely unrelated to the
probes of the GA, and so we have in these plots quite striking independent
corroboration of the GA. Furthermore, the tendency of the Mg~{\sc II}
absorbers in general and the quasars to share common paths and voids is
apparent, especially so in Fig.~\ref{fig:MgII_and_dr16_quasars_imag_19_5}.

Note that there is a density boundary in the distribution of the DR16Q
quasars: in roughly the lower third of the plots the density of the quasars
is lower than above. This artefact, however, is well separated from the GA
and does not affect our visual assessment.

\begin{figure}
    \centering
    \hspace*{-1.0cm}
    \includegraphics[scale=0.23]{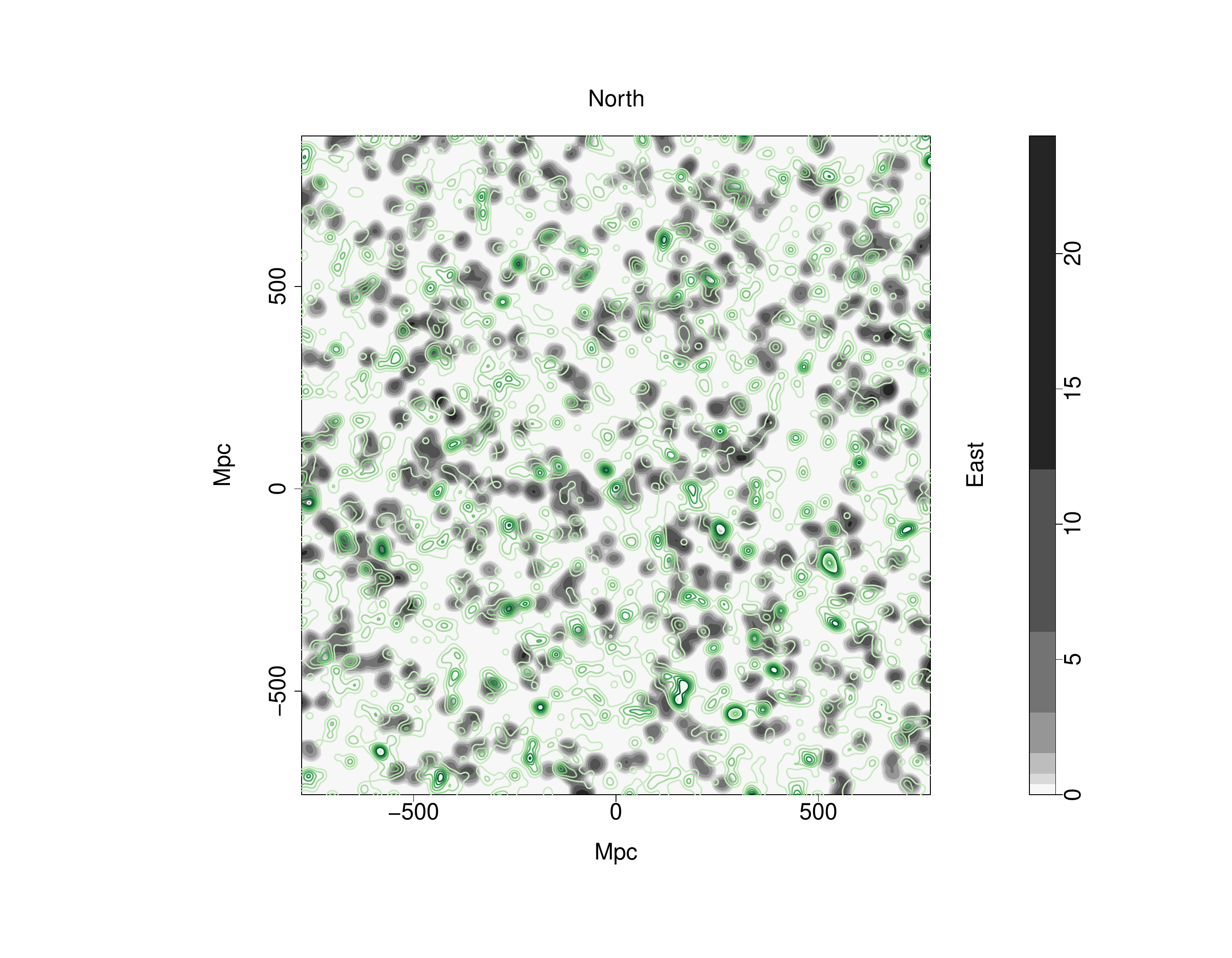}
    \caption{Density distribution of the flat-fielded Mg~{\sc II} absorbers
      in the redshift slice $z = 0.802 \pm 0.060$ represented by the grey
      contours which have been smoothed using a Gaussian kernel of $\sigma =
      11$~Mpc and increase by a factor of two.  Green contours represent the
      DESI clusters, of all richnesses, in the same redshift slice as the
      Mg~{\sc II} absorbers, smoothed using a Gaussian kernel of $\sigma =
      11$~Mpc and increasing by a factor of two. The GA can be seen
      stretching across $\sim 1$~Gpc in the centre of the figure (at
      tangent-plane $y$-coordinate $\sim 0$~Mpc).  There are no compelling
      connections between the DESI clusters and the Mg~{\sc II} absorbers.}
    \label{fig:MgII_and_DESI}
\end{figure}

We continue with the DESI clusters, again selected for the same redshift
interval as the GA --- Fig.~\ref{fig:MgII_and_DESI}. Note that the redshifts
for the DESI clusters are photometric, with redshift errors $\sim 0.024$ at
$z \sim 0.9$ \citep{Zou2021}. (In contrast, we might expect the redshift
errors for the quasars to be $\sim 0.003$.)

There is no compelling association of the DESI clusters and the GA, although
there is perhaps a hint on the RHS. Possibly the substantial errors in the
photometric redshifts are a factor in diluting any correspondence that might
exist.  An interesting feature in Fig.~\ref{fig:MgII_and_DESI} is, however,
the `cluster of clusters' in the centre of the GA, largely coinciding with
the central small gap in the Mg~{\sc II} absorbers of the GA. It could be a
large supercluster, with the SZ cluster B18, mentioned previously, as one of
its member clusters. We previously mentioned, in
Section~\ref{sec:observational_properties_GA}, that there appears to be a set
of strong Mg~{\sc II} absorbers enveloping a circular hole in the centre of
the GA. It seems likely that these enveloping strong absorbers and the
central hole are related to this putative supercluster.

\begin{figure}
    \centering
    \hspace*{-1.0cm}
    \includegraphics[scale=0.23]{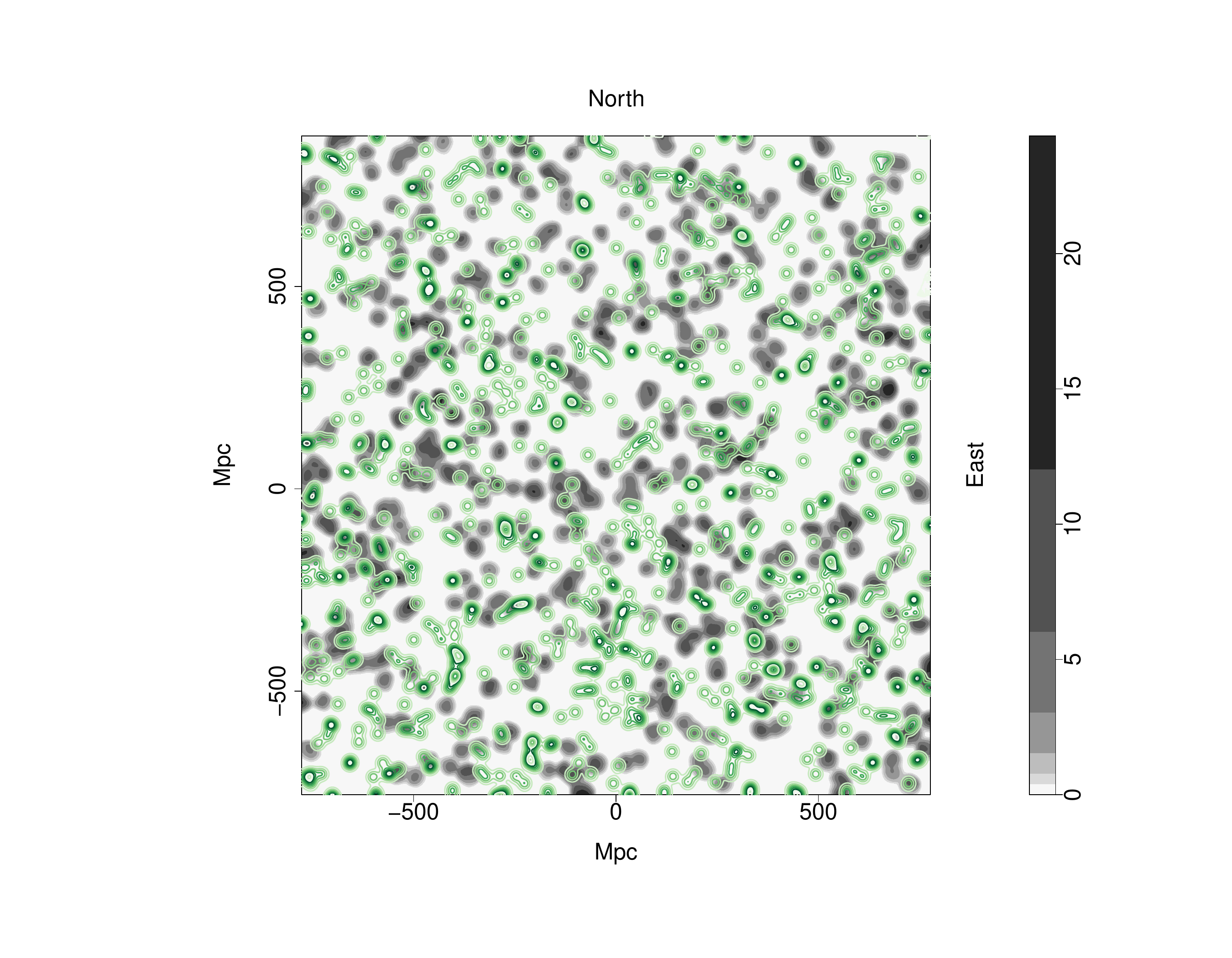}
    \caption{Density distribution of the flat-fielded Mg~{\sc II} absorbers
      in the redshift slice $z = 0.802 \pm 0.060$ represented by the grey
      contours which have been smoothed using a Gaussian kernel of $\sigma =
      11$~Mpc and increase by a factor of two.  Green contours represent the
      DESI clusters, with the richness limit $R \le 22.5$, in the same
      redshift slice as the Mg~{\sc II} absorbers, smoothed using a Gaussian
      kernel of $\sigma = 11$~Mpc and increasing by a factor of two. The GA
      can be seen stretching across $\sim 1$ Gpc in the centre of the figure
      (at tangent-plane $y$-coordinate $\sim 0$~Mpc). There are a few
      occurrences of the green contours following the grey contours,
      indicating that there might be some association of low richness
      clusters with Mg~{\sc II} absorbers.}
    \label{fig:MgII_and_DESI_rich_leq_22_5}
\end{figure}

The mean richness limit for the DESI clusters is $22.5$ \citep{Zou2021}.
Fig.~\ref{fig:MgII_and_DESI_rich_leq_22_5} shows the relationship between the
Mg~{\sc II} absorbers and DESI clusters with richness $R \le 22.5$.  It
suggests that there could be some association of the low-richness clusters
with the Mg~{\sc II} absorbers, both for the GA and in general.

\begin{figure}
    \centering
    \includegraphics[scale=0.20]{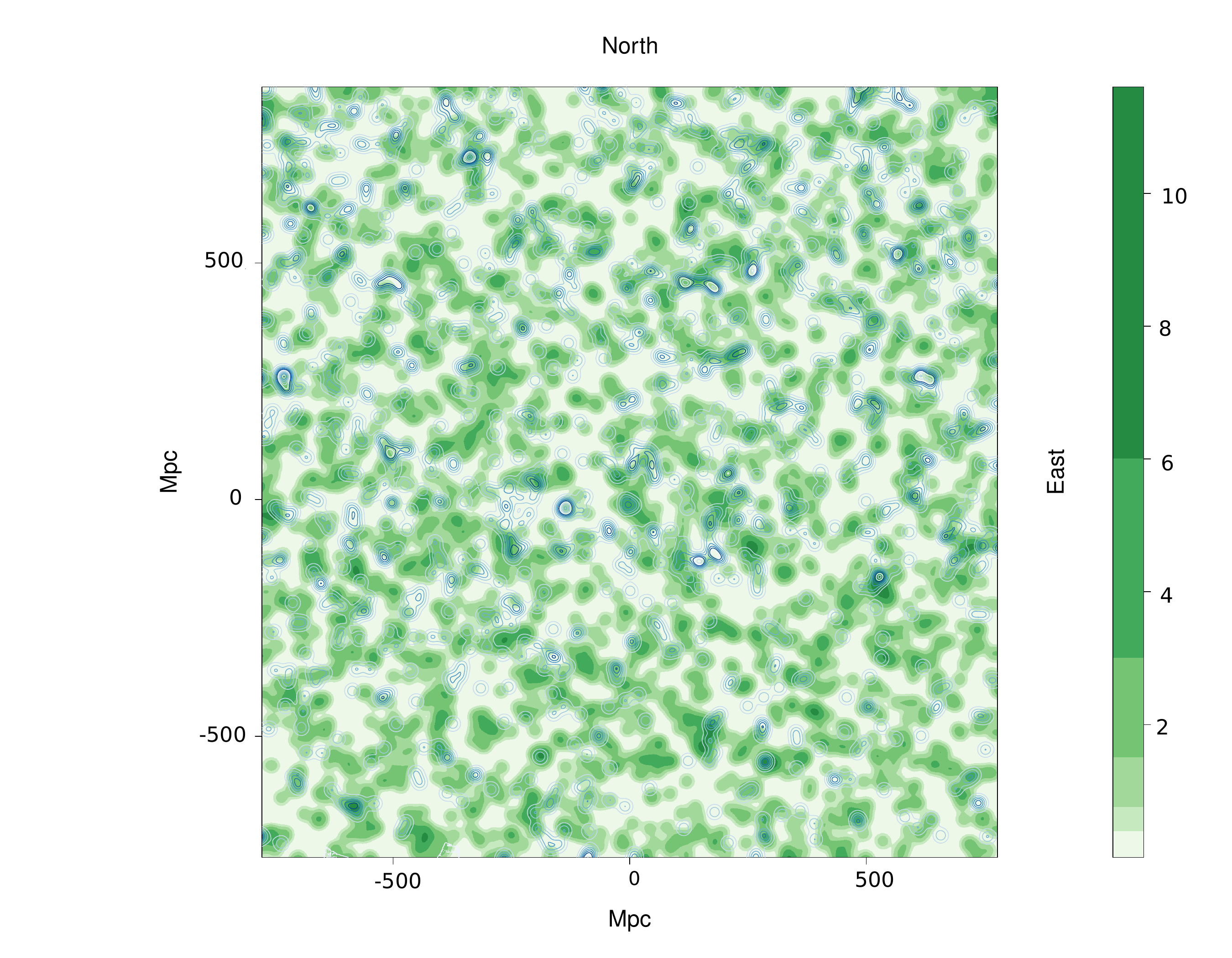}
    \caption{Density distribution of the DESI clusters in the redshift slice
      $z= 0.802 \pm 0.060$ represented by the green contours which have been
      smoothed using a Gaussian kernel of $\sigma = 11$~Mpc and increase by a
      factor of two. Blue contours represent the DR16Q quasars, with the
      magnitude limit $i \le 20.0$, in the same redshift slice as the DESI
      clusters, smoothed using a Gaussian kernel of $\sigma = 11$~Mpc and
      increasing by a factor of two. There are no compelling connections
      between the DR16Q quasars and the DESI clusters.}
    \label{fig:DESI_and_dr16_quasars}
\end{figure}

\begin{figure}
    \centering
    \includegraphics[scale=0.20]{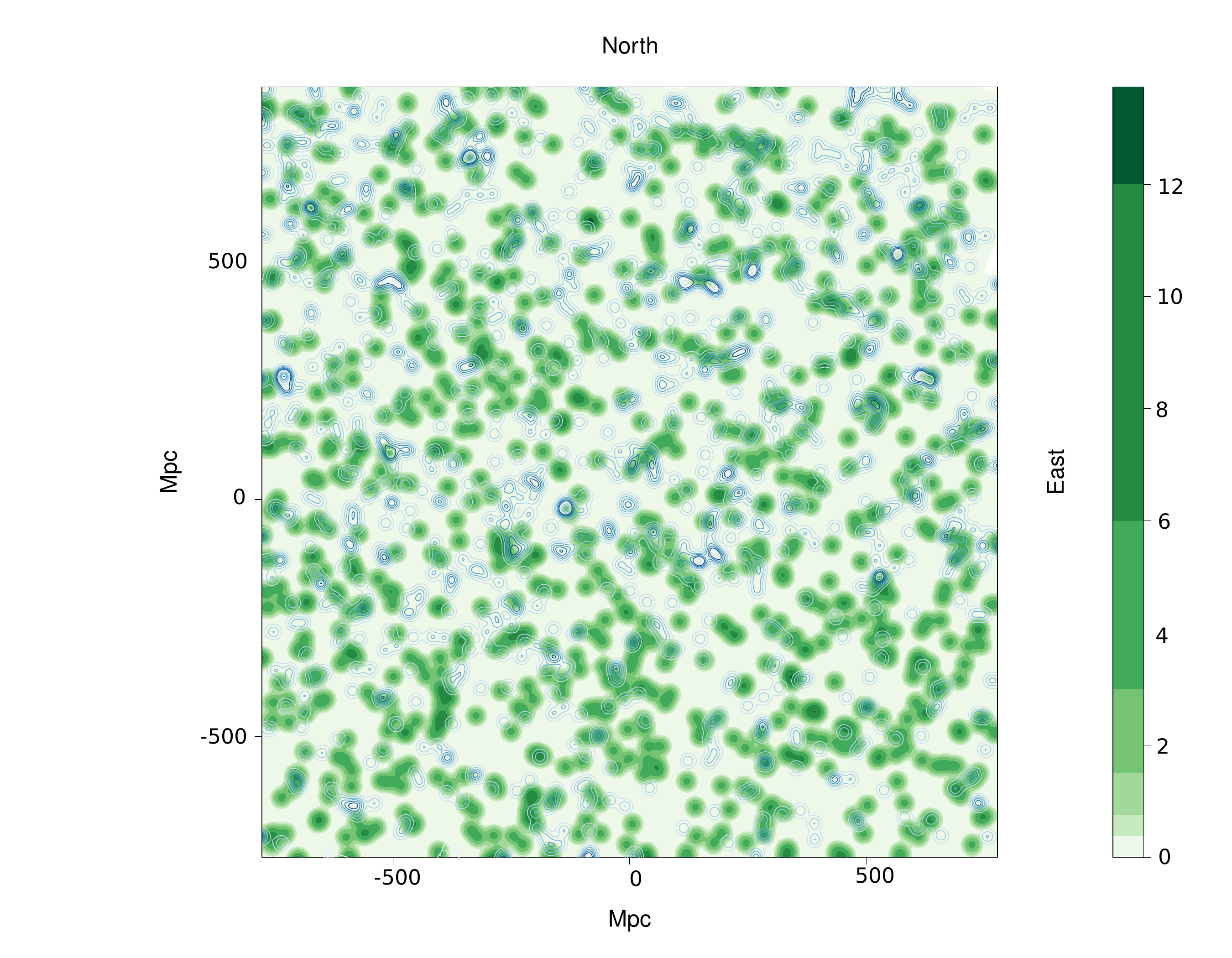}
    \caption{Density distribution of the DESI clusters in the redshift slice
      $z= 0.802 \pm 0.060$ represented by the green contours, increasing by a
      factor of two, which have been smoothed using a Gaussian kernel of
      $\sigma = 11$~Mpc and limited to show only those clusters with a
      richness $R \leq 22.5$. Blue contours represent the DR16Q quasars, with
      the magnitude limit $i \le 20$, in the same redshift slice as the DESI
      clusters, smoothed using a Gaussian kernel of $\sigma = 11$~Mpc and
      increasing by a factor of two. There are a few occurrences of the blue
      contours following the green contours, indicating that there might be
      some association of DR16Q quasars with low richness clusters.}
    \label{fig:DESI_and_dr16_quasars_rich_leq_22_5}
\end{figure}

Finally, we compare the the DESI clusters with the DR16Q quasars ---
Fig.~\ref{fig:DESI_and_dr16_quasars}.  As with the Mg~{\sc II} absorbers and
DESI clusters, there appears to be no compelling association. However, again,
the lowest-richness clusters suggest some association ---
Fig.~\ref{fig:DESI_and_dr16_quasars_rich_leq_22_5}.

From the independent corroboration above, we suggest that the GA, and the
Mg~{\sc II} absorbers in general, are associated with luminous quasars but
not strongly with DESI clusters.  However, there is potentially an
association of the Mg~{\sc II} absorbers and the quasars with the low
richness clusters.  More statistical details of the relationship between
Mg~{\sc II} absorbers, quasars and clusters will be investigated in our
future work.

\section{Discussions and Conclusions}

In this paper we have presented the discovery of the Giant Arc (GA): a
$\sim$~Gpc LSS at $z \sim 0.8$, mapped by Mg~{\sc II} absorption systems in
the spectra of background quasars. The GA forms a large crescent shape on the
sky that appears almost symmetrical.  However, deeper analysis reveals some
asymmetries in the GA, in the redshift and equivalent width (EW)
distributions. The GA spans $\sim 1$~Gpc on the sky and has a redshift depth
of $\sim 340$~Mpc (both proper sizes, present epoch) . Visually, we determine
the GA as a single unit, but using a Minimal Spanning Tree (MST) type
algorithm (see Section~\ref{subsect:MST}) it splits into two portions: a
large portion (GA-main) and a small portion (GA-sub).  We proposed in
Section~\ref{subsect:MST} that the two portions of the GA could in fact be
connected in reality, since potentially one more background probe could lead
to one more Mg~{\sc II} absorber that would connect the two portions.  On its
own, GA-main is a statistically-significant clustering of Mg~{\sc II}
absorbers, with a membership of 44 Mg~{\sc II} absorbers, an MST-overdensity
of $1.3 \pm 0.3$, and a mass excess of $1.8 \times 10^{18} M_\odot$. In these
respects, the GA is comparable to the Huge-LQG \citep{Clowes2013}.

Three different statistical tests were applied to the GA to assess the
significance of connectivity and clustering.  The results of each are
summarised here.  (i) The SLHC / CHMS method calculates the significance of
clustering between points of close proximity by comparing the volumes of the
CHMS for each structure to the CHMS of structures in randomly distributed
points in a cube. GA-main, containing 44 Mg~{\sc II} absorbers, has a
significance of $(4.5 \pm 0.6)\sigma$. GA-sub, containing 11 Mg~{\sc II}
absorbers, has a significance of $(2.1 \pm 0.9)\sigma$. Both GA-main and
GA-sub have the same MST-overdensity of $\delta \rho / \rho = 1.3 \pm 0.3$.
This fact could indicate, as we suspect, that both agglomerations are
connected in reality.  (ii) The CE test is a case-control $k$
nearest-neighbour algorithm that assesses the $p$-value of clustering in the
field within an unevenly distributed population. A process of zooming into
the GA field allows the GA to become increasingly dominant.  In this way, we
detect a $p$-value of $0.0027$ from the field seen in
Fig.~\ref{fig:CE_MgII_GA_zoom2}, equivalent to a significance of $3.0\sigma$.
Applying this process of zooming to lower redshift fields at the same sky
coordinates of the GA we do not detect any significant clustering.  We then
use our polygon approach that randomises points outside the GA while keeping
the visually selected absorbers contained within the GA the same. The CE test
detects similar $p$-value of clustering with the polygon approach, indicating
that the GA is the true, dominating feature causing significant clustering.
(iii) The PSA is a Fourier method of detecting clustering in the field on a
physical scale. We apply the 2D PSA to the `zoomed' GA field,
Fig.~\ref{fig:CE_MgII_GA_zoom2}, and find significant clustering at
$\lambda_c \sim 270\,\mathrm{Mpc}$ with a significance of $4.8\sigma$.  As
with the CE test, we use our polygon approach and detect similar significant
clustering scales. However, a small contribution from other absorbers in the
field is also detected. We do expect this given the power of the PSA test,
and it is clear that the GA is still the dominant contributor to the PSA
result.

Clearly, the analysis of the GA is after the event of its discovery, as is
unavoidable with unexpected discoveries in astrophysics and cosmology.  We
have applied several different approaches to mitigating any {\it post-hoc}
aspects of analysing the statistical significance of the GA after
discovery. We have performed techniques that aim to assess the GA
unbiassedly, such as the polygon approach, varying redshift slices, zooming
into the GA field, and randomised simulations. In the future we can apply the
same techniques used for the GA field to the whole of the Mg~{\sc II}
dataset. In addition, the Mg~{\sc II} dataset is quite complex, with features
that need careful attention: for example, the inhomogeneities of the Mg~{\sc
  II} absorbers are superimposed on the inhomogeneities of the quasars
(background probes) and of the survey. Finally, there are different Mg~{\sc
  II} databases available from different authors, each using different
detection processes. We intend eventually to produce our own databases of
Mg~{\sc II} detections that can be used consistently with past and future
quasar-survey data releases.

The GA is now amongst several other very large LSS discoveries with sizes
that exceed the theoretical upper-limit scale of homogeneity of
\citet{Yadav2010}. Potentially, there are other such significant structures
in the rest of the Mg~{\sc II} database. We discuss that there are challenges
in fairly characterising the population of structures due to the
inhomogeneities in the background probes (quasars). However, the challenges
can be managed with suitable care, allowing for the Mg~{\sc II} method of
studying LSS to be fully exploited.

In Table~\ref{tab:lss_list}, we listed some of the very large LSSs, and also
some of the reported CMB anomalies.  In standard cosmology we expect to find
evidence for a homogeneous and isotropic universe. However, the accumulated
set of LSS and CMB anomalies now seems sufficient to constitute a prima facie
challenge to the assumption of the Cosmological Principle (CP).  A single
anomaly, such as the GA on its own, could be expected in the standard
cosmological model. For example, \citet{Marinello2016} find that the Huge-LQG
\citep{Clowes2013}, a structure comparable in size to the GA, is, by itself
(there are others), compatible with the standard cosmological model. However,
\citet{Marinello2016} state that this is on the condition that only one
structure as large as the Huge-LQG is found in a field $\sim 5$ times the
sample survey, in this case, the DR7QSO quasar database for $1.2 \le z \le
1.6$.  Note that the GA is found in the combined footprint from DR7QSO and
DR12Q (the combined footprint being almost the same area as the individual
footprints), in a narrow redshift interval, so its challenge to the CP seems
likely to be exacerbated. Of course, the GA is now the fourth largest LSS, so
there are, at minimum, four LSSs comparable to the size of the Huge-LQG, plus
several other LSSs exceeding the scale of homogeneity. We suggest that there
is a need to explore other avenues within cosmology that could explain
multiple, very large LSSs.

We bring attention to the Sloan Great Wall (SGW) \citep{Gott2005}, which is a
large, wall-like filament in the relatively local universe. The SGW is $\sim
450$~Mpc in its longest dimension, which is $\sim 0.5$ times the length of
the GA. One can note some of the similarities between the SGW and the GA,
such as the general shape and comoving size --- they are both long,
filamentary and curved walls made up of galaxies and galaxy clusters ---, and
so perhaps also envision a LSS such as the GA as a precursor to the SGW. The
GA is at a redshift of $\sim 0.8$ which means we are seeing it when the
universe was only half its present age. Perhaps the SGW, at an earlier epoch,
initially looked more like the GA. At this point, these ideas are speculative
only, but experimenting with simulations (possibly even with alternative
cosmological models) could conceivably elucidate such hypothetical
connections between structures like the GA and the SGW.

\section*{Acknowledgements}

We thank Srinivasan Raghunathan for many helpful discussions, and we thank
Ilona S\"ochting for suggesting use of the Cuzick-Edwards test.

This paper has depended on SDSS data and on the R software.

We thank the replacement referee for careful reading and thoughtful comments.

\section*{Data availability statement}

The datasets were derived from sources in the public domain: {\small
  https://www.guangtunbenzhu.com/jhu-sdss-metal-absorber-catalog}.


\bsp	
\label{lastpage}
\end{document}